\documentclass[%
 aip,
 amsmath,amssymb,
preprint,%
]{revtex4-1}

\usepackage{graphicx}% Include figure files
\usepackage{dcolumn}% Align table columns on decimal point
\usepackage{bm}% bold math
\usepackage[utf8]{inputenc}
\usepackage[T1]{fontenc}
\usepackage{mathptmx}
\usepackage{makecell}
\usepackage{xcolor}

\begin{document}

% \preprint{AIP/123-QED}

\title[]{Developed and Quasi-Developed Macro-Scale Flow in Micro- and Mini-Channels with Arrays of Offset Strip Fins}
% \title[]{Macro-Scale Description of Flow and Heat Transfer in Micro- and Mini-Channels with Arrays of Offset Strip Fins subject to a uniform heat flux (PART I: Flow)}
% Force line breaks with \\

\author{A. Vangeffelen}
\email{arthur.vangeffelen@kuleuven.be.}
\author{G. Buckinx}%
\affiliation{ 
Department of Mechanical Engineering, KU Leuven, Celestijnenlaan 300A, 3001 Leuven, Belgium%\\This line break forced with \textbackslash\textbackslash
}%
\affiliation{%
VITO, Boeretang 200, 2400 Mol, Belgium%\\This line break forced% with \\
}%
\affiliation{%
EnergyVille, Thor Park, 3600 Genk, Belgium%\\This line break forced% with \\
}%

\author{C. De Servi}
\affiliation{%
VITO, Boeretang 200, 2400 Mol, Belgium%\\This line break forced% with \\
}%
\affiliation{%
EnergyVille, Thor Park, 3600 Genk, Belgium%\\This line break forced% with \\
}%

\author{M. R. Vetrano}
\affiliation{ 
Department of Mechanical Engineering, KU Leuven, Celestijnenlaan 300A, 3001 Leuven, Belgium%\\This line break forced with \textbackslash\textbackslash
}%
\affiliation{%
EnergyVille, Thor Park, 3600 Genk, Belgium%\\This line break forced% with \\
}%

\author{M. Baelmans}
\affiliation{ 
Department of Mechanical Engineering, KU Leuven, Celestijnenlaan 300A, 3001 Leuven, Belgium%\\This line break forced with \textbackslash\textbackslash
}%
\affiliation{%
EnergyVille, Thor Park, 3600 Genk, Belgium%\\This line break forced% with \\
}%

\date{\today}% It is always \today, today,

\begin{abstract}
We investigate to what degree the steady laminar flow in typical micro- and mini-channels with offset strip fin arrays can be described as developed on a macro-scale level, in the presence of channel entrance and side-wall effects. 
% This macro-scale description is based on a double volume-averaging operation of the flow variables and corresponds to a developed friction factor correlation. 
Hereto, the extent of the developed and quasi-developed flow regions in such channels is determined through large-scale numerical flow simulations. 
It is observed that the onset point of developed flow increases linearly with the Reynolds number and channel width, but remains small relative to the total channel length. 
Further, we find that the local macro-scale pressure gradient and closure force for the (double) volume-averaged Navier-Stokes equations are adequately modeled by a developed friction factor correlation, as typical discrepancies are below 15\% in both the developed and developing flow region. 
We show that these findings can be attributed to the eigenvalues and mode amplitudes which characterize the quasi-developed flow in the entrance region of the channel. 
Finally, we discuss the influence of the channel side walls on the flow periodicity, the mass flow rate, as well as the macro-scale velocity profile, which we capture by a displacement factor and slip length coefficient. 
Our findings are supported by extensive numerical data for fin height-to-length ratios up to 1, fin pitch-to-length ratios up to 0.5, and channel aspect ratios between 1/5 and 1/17, covering Reynolds numbers from 28 to 1224. 
\\ % and macro-scale closure force

\end{abstract}

\maketitle

\section{\label{sec:intro}Introduction}

%% Context: Modelling pressure drop and heat transfer in mini- and micro-channels with large arrays of solid structures
% Macro-scale description based on double volume-average (VAT = periodic eqs.)
% 	 Exact and physically meaningful
% Previous work: Periodically developed flow and heat transfer regime
% 	 Correspond to the exact closure models in the macro-scale framework

% However: Entrance and exit effects, and side wall influence in full channel
%  Where are the closure models valid? 
%  Can they be extended to the side-wall region of a full channel? 
%  How large are the discrepancies in the in- and outlet region of the channel? 
%  What is the effect of the geometry, Reynolds number and Prandtl number?

% How: Numerical simulations of the developing flow and heat transfer regime through entire arrays of periodic solid structures, in a macro-scale framework

%% offset strip fin micro- and mini-channels
Micro- and mini-channels with arrays of periodic fins have been increasingly applied in highly compact heat transfer devices over the last twenty years (Refs.~\onlinecite{kandlikar2005heat,khan2006role,izci2015effect,yang2017heatpin}). 
In particular, micro- and mini-channels with arrays of offset strip fins are frequently employed in high-power-density heat transfer devices. 
Their applications are the cooling of microelectronics (Refs.~\onlinecite{bapat2006thermohydraulic, yang2007advanced, hong2009three}), heat recuperation in compact gas turbines (Refs.~\onlinecite{do2016experimental,nagasaki2003conceptual}), refrigeration and liquefaction in cryogenic systems (Refs.~\onlinecite{yang2017heat,jiang2019thermal}), and air heating in solar collectors (Refs.~\onlinecite{yang2014design,pottler1999optimized}). 

Micro- and mini-channels are defined by their smallest dimensions, which lie between 10 $\mu$m to 200 $\mu$m, and 200 $\mu$m to 3 mm, respectively (Ref.~\onlinecite{kandlikar2005heat}). 
Furthermore, in micro- and mini-channels with offset strip fin arrays, the fin height is commonly smaller than the fin length (Refs.~\onlinecite{tuckerman1981high, bapat2006thermohydraulic, yang2007advanced, hong2009three,do2016experimental,nagasaki2003conceptual,yang2017heat,jiang2019thermal,yang2014design,pottler1999optimized}). 
Due to these small channel and fin dimensions, the flow inside offset strip fin micro- and mini-channels typically remains in a laminar and steady regime.
This laminar steady regime is characterized by a low to moderate Reynolds number between 10 and 500 (Refs.~\onlinecite{tuckerman1981high,bapat2006thermohydraulic,yang2007advanced,hong2009three,do2016experimental,nagasaki2003conceptual,yang2017heat,jiang2019thermal,yang2014design,pottler1999optimized,zargartalebi2020impact}), as we underlined in our previous work (Refs.~\onlinecite{vangeffelen2021friction,vangeffelen2022nusselt}).

%% Modelling pressure drop and heat transfer
To assess the hydraulic performance of micro- and mini-channels with fin arrays, the relationship between the pressure drop over the channel and the flow rate through the channel is often correlated by means of numerical flow simulations (Refs.~\onlinecite{kim2011correlations,yang2014general}). %  and heat transfer
For such flow simulations, only a single unit cell of the fin array is usually considered (Refs.~\onlinecite{liang2022fluid,odele2022performance}). 
This significantly reduces the required computational resources, in contrast to Direct Numerical Simulation (DNS) of the detailed flow throughout the entire channel (Ref.~\onlinecite{kim2010thermoflow}). 
In general, two different approaches, or theoretical frameworks, can be distinguished to characterize the flow through a unit-cell simulation. % unit cell approaches

In the first approach, it is assumed that the flow is periodically developed, and thus similar in every unit cell of the fin array.
As such, the pressure drop over the unit cell for a given flow rate is determined by solving the periodically developed flow equations (Ref.~\onlinecite{patankar1977fully}).
The latter govern the periodic components of the velocity and pressure fields on the unit cell (Refs.~\onlinecite{krishnan2008simulation,kim2010thermoflow,alshare2010modeling}). 

In the second approach, the fin array is treated as a porous medium, so that one can rely on the volume-averaging technique (VAT) for porous media (Refs.~\onlinecite{whitaker1996forchheimer,quintard1997two}) to obtain the volume-averaged or \textit{macro-scale} pressure gradient in the channel for a given flow rate. 
Then, the relationship between the macro-scale pressure gradient and the flow rate is expressed by an (apparent) permeability tensor. 
This permeability tensor is the solution of a so-called closure problem, which governs the (non-averaged) deviation components of the velocity and pressure fields on the unit cell (Refs.~\onlinecite{saito2006correlation,raju2007porous}). 
As the most commonly adopted closure problem (Refs.~\onlinecite{whitaker1996forchheimer,quintard1997two}) has the same mathematical form as the periodically developed flow equations (Ref.~\onlinecite{patankar1977fully}), a distinction between both approaches is in practice not always made (Refs.~\onlinecite{kim2000local,nakayama2004numerical}). 

Nevertheless, from a theoretical viewpoint, both approaches are only consistent and equivalent when a specific type of volume-averaging technique is used. 
As Buckinx and Baelmans (Refs.~\onlinecite{buckinx2015multi,buckinx2015macro,buckinx2016macro}) have shown, an exact macro-scale description of periodically developed flow requires that the macro-scale velocity and pressure are defined through a double volume-averaging operation, which was originally introduced by Quintard and Whitaker (Refs.~\onlinecite{quintard1994transport1,quintard1994transport2,quintard1994transport3,quintard1994transport4,quintard1997two,davit2017technical}). 
With this double volume-averaging technique, also a physically meaningful macro-scale description is achieved.
First, it results in a spatially constant macro-scale pressure gradient in the developed regime, which agrees with the actual pressure drop over each fin unit.
Secondly, it leads to a spatially constant macro-scale velocity, which corresponds to the actual flow rate through the unit cell. 
Moreover, it allows us to represent the \textit{developed} macro-scale pressure gradient, as well as the closure force exerted by the solid fins on the flow, by means of a spatially constant permeability tensor.
Notably, the same double volume-averaging technique is also required to construct an exact and physically meaningful macro-scale description of the periodically developed heat transfer regimes (Refs.~\onlinecite{buckinx2015macro,buckinx2016macro,vangeffelen2022nusselt}).

%We remark that, as an alternative to the volume-averaging methods, also macro-scale flow models have been obtained through homogenization of the Navier-Stokes equations on the basis of two-scale asymptotic expansions (Refs.~\onlinecite{sanchez-palencia_fluid_1980,allaire_homogenization_1991}). 
%However, the relationship between these asymptotic expansions and particular solutions to the Navier-Stokes equations like the (quasi-) periodically flow equations in the macro-scale framework of Buckinx have not yet been explored. 

% In our previous work (Ref.~\onlinecite{vangeffelen2021friction}), we have analysed the periodically developed flow regime in micro- and mini-channels with offset strip fin arrays. 
In our previous works (Refs.~\onlinecite{vangeffelen2021friction,vangeffelen2022nusselt}), we have analyzed the periodically developed flow and heat transfer regime in micro- and mini-channels with offset strip fin arrays. 
In particular, in (Ref.~\onlinecite{vangeffelen2021friction}), we have correlated the developed macro-scale pressure gradient in the form of a dimensionless friction factor, as a function of the Reynolds number and the geometrical parameters of the offset strip fin array.
%corresponding to an apparent permeability tensor of the offset strip fin unit cell domain. 
However, it is still unknown after which distance from the channel inlet the flow can be regarded as periodically developed in such channels.
Therefore, it is unclear to what extent the latter macro-scale description is valid for common applications of micro- and mini-channels with arrays of offset strip fins.
% In particular/Additionally, up to now, 
More precisely, it still needs to be investigated how accurately the developed friction factor correlation from (Ref.~\onlinecite{vangeffelen2021friction}) can represent the macro-scale pressure gradient (or closure force) in the entrance region, where the flow is developing, and therefore model the pressure drop over the entire offset strip fin array. 

% In the literature, few studies exist on the flow development in micro- and mini-channels with an array of offset strip fins. 
In the literature, flow development has almost exclusively been studied in channels of a constant cross section, hence without solid fins, and mainly for two-dimensional laminar channel flows (Refs.~\onlinecite{schiller1922entwicklung,chen1973flow,langhaar1942steady,sparrow1964flow,schlichting1934laminar,kapila1973entry,brandt1969asymptotic,wilson1969development,sadri2002accurate,patankar1983calculation,ferreira2021hydrodynamic}).
Hereto, various approaches have been explored to describe developing flow. 
A first approach is the integral method, according to which the channel flow is divided into two regions. 
In the region near the channel wall, the flow is assumed to form a developing boundary layer, whereas, in the channel center, the flow is considered inviscid (Refs.~\onlinecite{schiller1922entwicklung,chen1973flow}).
In particular, Chen et al. (Ref.~\onlinecite{chen1973flow}) used the integral method to determine the flow development length in circular pipes and parallel-plate channels. 
Their theoretical analysis predicted a linear scaling of the dimensionless flow development length with the Reynolds number in the laminar regime, for Reynolds numbers below fifty. 

A second approach relies on the linearization of the inertial terms in the Navier-Stokes equations (Refs.~\onlinecite{langhaar1942steady,sparrow1964flow}). 
The linearized flow equations give rise to a type of problem that can be solved by means of Bessel functions. 
% so that the developing velocity profile in the entrance region can be determined by Bessel functions (Refs.~\onlinecite{langhaar1942steady,sparrow1964flow}). 
That way, the flow development length and the incremental pressure drop in the channel entrance region can be approximated. 
The application of the linearization approach is restricted to simple channel geometries such as parallel-plate channels and circular pipes, in which the transversal flow has a negligible influence on the overall velocity distribution. 

As a third approach, the developing flow can be described through matched asymptotic expansions (Refs.~\onlinecite{schlichting1934laminar,kapila1973entry}). 
This means that first an asymptotic approximation for the flow field near the channel inlet is obtained, by solving the perturbed boundary-layer flow equations. 
This approximation is then matched to a second asymptotic approximation for the flow further downstream of the inlet, which is obtained as a perturbation from the fully-developed flow equations. 
%The developing flow problem is solved by matching both asymptotic expansions at an intermediate section. 
% a series expansion (Refs.~\onlinecite{schlichting1934laminar,kapila1973entry}). 
%For this method, a perturbation of the boundary layer flow model is assumed near the channel inlet, while, far downstream of the inlet, a perturbation of the fully developed Poisseuille flow is assumed. 
Notably, the asymptotic velocity modes which characterize the quasi-developed flow further downstream of the inlet have also been analyzed on their own, at least for laminar flows in parallel-plate channels at Reynolds numbers up to 2200 (Refs.~\onlinecite{brandt1969asymptotic,wilson1969development,sadri2002accurate}). 
These modes satisfy an eigenvalue problem similar to the Orr-Sommerfeld differential equation (Ref.~\onlinecite{lin1955theory}), which is typically solved using an initial type method or a spectral method (Refs.~\onlinecite{nachtsheim1964initial,nagy2019stabilization,canuto2012spectral}). 

Finally, numerical approaches have been adopted to solve the developing flow by direct numerical simulations of the complete Navier-Stokes equations (Ref.~\onlinecite{sadri1997channel}). 
As a matter of fact, for three-dimensional flows in channels without solid structures, such numerical approaches (Refs.~\onlinecite{patankar1983calculation,ferreira2021hydrodynamic,gessner1981numerical}) are still the only means to obtain the flow development length and the incremental pressure drop, for both the laminar and turbulent regime. 

Up to the present, only a few studies have been conducted on the flow development in channels with arrays of periodic fins (Ref.~\onlinecite{abd1986flow,shams2018numerical,joshi1987heat,mochizuki1988flow,dejong1997experimental}) or other solid structures such as baffles (Ref.~\onlinecite{mousavi2006heat}) and tube bundles (Ref.~\onlinecite{morrison1997flow}). 
However, such studies mainly investigate transient flow conditions where vortex shedding takes place in the wakes of the solid structures. 
% Is this really the main conclusion/motivation behind all the studies you just mentioned? Let us discuss about this in person further ..
% However, most of the work ... like baffles and tube bundles. 

A first theoretical and numerical study of laminar flow development in channels with arrays of in-line square cylinders has recently been presented by Buckinx (Ref.~\onlinecite{buckinx2022arxiv}).
This recent work introduces a macro-scale description of the developing flow upstream of the periodically developed flow region.
In this region of so-called quasi-periodically developed flow, the flow velocity is determined by a single mode that decays exponentially in the streamwise direction, with a streamwise periodic amplitude.
As such, it was shown possible to derive a local closure problem for the apparent permeability tensor which is exact for most of the developing flow region. 

Although no prior work has considered the flow development in micro- and mini-channels with offset strip fins, some experimental data for conventional channels with offset strip fins has been provided by Dong et al. (Ref.~\onlinecite{dong2007air}).
Dong et al. measured the pressure profile in 16 offset strip fin channel geometries, each having a total channel length between 5 and 14 fin lengths.
From their experimental data, the influence of flow development on the total pressure drop over the channel was captured by a friction factor correlation which includes the total channel length.
Nevertheless, due to the relatively large height of conventional channels, their data applies only to the transitional flow regime.
In the transitional flow regime, the flow exhibits an onset point to vortex shedding in the wakes of the most downstream fins, which travels upstream in the channel as the Reynolds number is increased (Refs.~\onlinecite{joshi1987heat,mochizuki1988flow,dejong1997experimental}). 
Consequently, the observations of Dong et al. have limited relevance for micro- and mini-channel applications in which steady laminar flow prevails. \\

%%% more background and literature
% For micro- and mini-channels without solid structures, the flow and heat transfer development has been investigated more extensively in the literature (ref: Schlichting, Langhaar, Lee, Sadri, ...)
% periodically developed heat transfer is expected to occur after a short development length from the channel inlet in the specified applications(Refs.~\onlinecite{lee2005investigation,lee2006thermally}). 

%% Research objective
In this work, we aim to investigate to what degree the flow in typical micro- and mini-channels with offset strip fins can be described as developed on a macro-scale level.
To that end, we will quantify the onset point of developed macro-scale flow in such offset strip fin channels.
In addition, we will quantify the extent of the region of quasi-developed macro-scale flow, and assess the accuracy of the macro-scale description of (Ref.~\onlinecite{buckinx2015macro}) in this region.
More specifically, we will determine the actual macro-scale closure force and actual macro-scale pressure gradient over the entire fin array to verify the accuracy of the developed friction factor correlation from our preceding work (Ref.~\onlinecite{vangeffelen2021friction}).
Lastly, we aim to investigate the influence of the channel side walls on the mass flow rate, the macro-scale velocity profile, as well as the macro-scale closure force in the developed flow region, as this influence has not been taken into account in the former friction factor correlation.
For these purposes, we rely on DNS of the flow in several micro- and mini-channel geometries with Reynolds numbers ranging from 28 to 1224. 
From these DNS results, the macro-scale variables are directly obtained by explicit spatial averaging (or filtering) with a discretized double volume-averaging operator.

%% Outline
The remainder of this work is structured as follows. 
In Section \ref{sec:method} the channel geometry and flow equations for DNS are discussed, together with the numerical procedure. 
The onset of the developed macro-scale flow regime and the accuracy of the developed friction factor correlation are discussed in Section \ref{sec:onsetMS}. 
In Section \ref{sec:onsetquasi}, the onset of the quasi-developed macro-scale flow and its modes are determined. 
Finally, in Section \ref{sec:sidewall}, we examine the influence of the side-wall region in offset strip fin micro- and mini-channels. \\

% \textcolor{red}{
% \textbf{Major Remark}\\
% Although I understand why we have used a Reynolds number based on the periodicity/fin length $l$ in our correlations for the friction factor (as it is the reference length over which the pressure drop occurs), it is very misleading to characterize the flow regime through the channel itself.
% I strongly recommend you characterize the flow regime based on the Reynolds number based on twice the channel height: $Re_{b} = \rho_f u_b (2h)/\mu_f$ at $h/l=0.12$, so that $Re_l=600$ should be replaced by $Re_{b} = 600 \times 0.12 \times 2=144$, like I did in my work.
% You can also use both reference lengths and thus two different Reynolds number everywhere, if you explain why you mention their difference in meaning.
% Further, I believe that it the low $h/l$ ratios and thus relatively low Reynolds number based on the height of the channel are responsible for the minor importance of flow development (next to the porosity etc.).
% }

\newpage
\clearpage

\section{\label{sec:method}Channel geometry and governing equations}

\subsection{Geometry}

\begin{figure*}[h!]
\includegraphics[scale = 0.75]{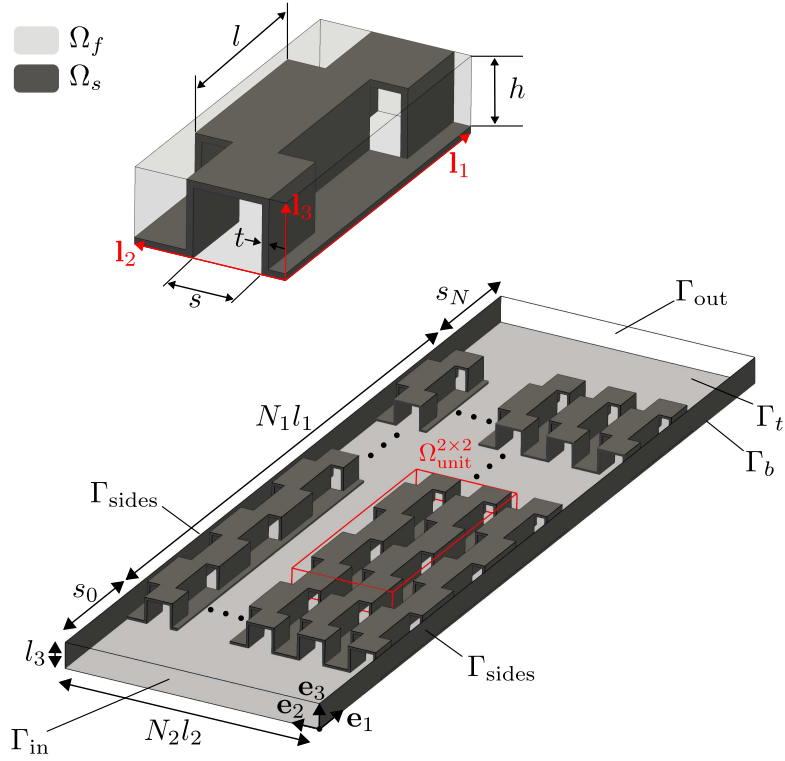}
\caption{\label{fig:geometrydns} Offset strip fin channel and unit cell domain}
\end{figure*}

%In order to determine the onset points of (quasi-) developed macro-scale flow, and to assess the accuracy of the developed friction factor correlation, a direct numerical simulation (DNS) of the flow through an entire array of offset strip fins is performed for a wide range of Reynolds numbers and geometrical parameters. 
A schematic of the three-dimensional channel domain and its fin array used for the direct numerical flow simulations is shown in Figure \ref{fig:geometrydns}. 
The geometrical parameters of a single unit cell of the fin array are the fin length $l$, the fin height $h$, the lateral fin pitch $s$ and the fin thickness $t$.
The porosity of the unit cell equals $\epsilon_{f} = hs/ \left[ \left( h+t \right) \left( s+t \right) \right]$. 
With respect to the normalized Cartesian vector basis $\{ \boldsymbol{e}_{j} \}_{j=1,2,3}$ in Figure \ref{fig:geometrydns}, the unit cell $\Omega_{\text{unit}}$ is spanned by the three lattice vectors $\boldsymbol{l}_{1} = l_{1} \boldsymbol{e}_{1} = 2l \boldsymbol{e}_{1}$, $\boldsymbol{l}_{2} = l_{2} \boldsymbol{e}_{2} = 2(s+t) \boldsymbol{e}_{2}$ and $\boldsymbol{l}_{3} = l_{3} \boldsymbol{e}_{3} = (h+t) \boldsymbol{e}_{3}$. 
The offset strip fin array in the channel consists of $N_1$ unit cells along the main flow direction and $N_2$ unit cells along the lateral direction. 
For all the cases considered in this work, the values $N_1=20$ and $N_2 = \{5, 10, 13, 15, 17 \}$ have been selected. 
% , relevant to micro- and mini-channel applications (Refs.~\onlinecite{bapat2006thermohydraulic, yang2007advanced, hong2009three,do2016experimental,nagasaki2003conceptual,yang2017heat,jiang2019thermal,yang2014design,pottler1999optimized}). 
As will be discussed in more detail in Sections \ref{sec:onsetMS} and \ref{sec:sidewall}, this ensures that a periodically developed flow region is established in the channel core, so that the onset of the developed macro-scale flow can be effectively determined. 
Moreover, given the selected values of $N_{1}$ and $N_{2}$, the developed flow in the channel core will not only be spatially periodic along the main flow direction but also along the transversal direction. 

% , relevant to micro- and mini-channel applications
In front and after the fin array, an inlet and outlet region without fins are provided with a length of $s_0$ and $s_N$, respectively. 
For the outlet region, a length $s_N = 2.5 l_{1}$ has been chosen, based on our observation that for $s_N \in ( 2.5 l_{1}, 10 l_{1} )$ the flow velocity fields inside the fin array remain practically identical so that the onset of periodically developed flow is not affected by the value of $s_N$. 
% the onset of the periodically developed region nor the position near the outlet where the flow in the fin array loses its periodicity. 
On the other hand, the selected value of the inlet length $s_0 = l_{1}$ does affect the flow development. %, as discussed further in more detail. 
% , as well as the selected inlet boundary condition,
%Nevertheless, the included channel geometries in this work allow to indicate the significance of the developed macro-scale flow region in offset strip fin micro- and mini-channels for various Reynolds numbers and geometrical parameters. 
% As a result, the total offset strip fin channel domain is spanned by $\boldsymbol{L}_{1} = L_{1} \boldsymbol{e}_{1} = (s_{0} + N_1 l_{1} + s_{N}) \boldsymbol{e}_{1}$, $\boldsymbol{L}_{2} = L_{2} \boldsymbol{e}_{2} = N_2 l_{2} \boldsymbol{e}_{2}$ and $\boldsymbol{L}_{3} = L_{3} \boldsymbol{e}_{3} = l_{3} \boldsymbol{e}_{3}$. 
In summary, the entire channel domain has a total length of $L_{1} = (s_{0} + N_1 l_{1} + s_{N})$, while the total width and height equal $L_{2} = N_2 l_{2}$ and $L_{3} = l_{3}$, respectively. 
We remark that with the fin length $l$ as the reference length, the channel geometry is fully specified by the non-dimensional geometrical parameters $h/l$, $s/l$, $t/l$, $s_0$, $s_N$, $N_{1}$, and $N_{2}$. 

The channel domain $\Omega$ is subdivided in a fluid domain $\Omega_{f}$ and a solid domain $\Omega_{s}$, which are separated by a fluid-solid interface $\Gamma_{fs}$. 
The exterior boundary of the channel domain $\Gamma = \partial \Omega$ has a unit normal vector $\boldsymbol{n}$ which points outward of the channel domain $\Omega$. 
This boundary $\Gamma$ consists of six planes, which represent the inlet $\Gamma_{\text{in}}$, the outlet $\Gamma_{\text{out}}$, the top wall $\Gamma_{t}$, the bottom wall $\Gamma_{b}$, and the side walls $\Gamma_{\text{sides}}$ of the channel. 
%Here, $\Gamma_{t}$, $\Gamma_{b}$ and $\Gamma_{\text{sides}}$ represent the solid walls which confine the channel. \\

\subsection{\label{sec:eqs}Steady channel flow equations}

The (steady) velocity field $\boldsymbol{u}_f$ and pressure field $p_{f}$ inside the offset strip fin channel are 
the numerical solution of the steady Navier-Stokes equations for an incompressible Newtonian fluid if the gravitational force or any other body forces are absent, 
% numerically obtained by solving the time-dependent Navier-Stokes equations for an incompressible Newtonian fluid, 
\begin{equation}
\begin{aligned}
  \nabla \cdot \boldsymbol{u}_{f} &= 0 & \text{in } \Omega_{f},  \\
  \rho_{f} \nabla \cdot \left( \boldsymbol{u}_{f} \boldsymbol{u}_{f} \right) &= - \mathrm{\nabla{p}}_{f} + \mu_{f} \nabla^2 \boldsymbol{u}_{f} & \text{in } \Omega_{f}, \\
  % \rho_{f} \frac{\partial \boldsymbol{u}_{f}}{\partial t} + \rho_{f} \nabla \cdot \left( \boldsymbol{u}_{f} \boldsymbol{u}_{f} \right) &= - \mathrm{\nabla{p}}_{f} + \mu_{f} \nabla^2 \boldsymbol{u}_{f} & \text{in } \Omega_{f}, \\
%   \boldsymbol{u}_{s} &= 0 & \text{in } \Omega_{s}, 
\end{aligned}
\label{eq:NavierStokes}
\end{equation}
where
\begin{equation}
\begin{aligned}
  \boldsymbol{u}_{f} \left( \boldsymbol{x} \right) &= \boldsymbol{u}_{\text{in}} \left( \boldsymbol{x} \right) & &\text{in } \Gamma_{\text{in}}, \\ %  = 36 u_{b} \frac{ x_{2} \left( L_{2} - x_{2} \right) x_{3} \left( L_{3} - x_{3} \right)}{L_{2}^{2} L_{3}^{2}} \boldsymbol{e}_{1}
  p_{f} \left( \boldsymbol{x} \right) &= 0 & &\text{in } \Gamma_{\text{out}}, \\
  \boldsymbol{u}_{f} \left( \boldsymbol{x} \right) &= 0 & &\text{in } \Gamma_{fs} \cup \Gamma_{t} \cup \Gamma_{b} \cup \Gamma_{\text{sides}}. 
  % \boldsymbol{u}_{f} \left( \boldsymbol{x} \text{,} t \right) &= \boldsymbol{u}_{\text{in}} \left( \boldsymbol{x} \right) & &\text{in } \Gamma_{\text{in}}, \\ %  = 36 u_{b} \frac{ x_{2} \left( L_{2} - x_{2} \right) x_{3} \left( L_{3} - x_{3} \right)}{L_{2}^{2} L_{3}^{2}} \boldsymbol{e}_{1}
  % p_{f} \left( \boldsymbol{x} \text{,} t \right) &= 0 & &\text{in } \Gamma_{\text{out}}, \\
  % \boldsymbol{u}_{f} \left( \boldsymbol{x} \text{,} t \right) &= 0 & &\text{in } \Gamma_{fs} \cup \Gamma_{t} \cup \Gamma_{b} \cup \Gamma_{\text{sides}}, \\
  % \boldsymbol{u}_{f} \left( \boldsymbol{x} \text{,} 0 \right) &= 0 & &\text{in } \Omega_{f}, \\
  % p_{f} \left( \boldsymbol{x} \text{,} 0 \right) &= 0 & &\text{in } \Omega_{f}. 
\end{aligned}
\label{eq:BoundaryConditions}
\end{equation}
Here, the fluid density $\rho_f$ and dynamic viscosity $\mu_f$ are assumed to be constants. 
At the inlet $\Gamma_{\text{in}}$, a parabolic velocity profile $\boldsymbol{u}_{\text{in}} = 36 u_{b} x_{2} \left( L_{2} - x_{2} \right) x_{3} \left( L_{3} - x_{3} \right) / \left( L_{2}^{2} L_{3}^{2} \right) \boldsymbol{e}_{1}$ has been imposed, such that the bulk average velocity is given by $u_{b} = - \int_{\Gamma_{\text{in}}} \boldsymbol{n} \cdot \boldsymbol{u}_{\text{in}} d\Gamma / \left( L_{2} L_{3} \right)$.
%, where the unit normal vector $\boldsymbol{n}$ is defined such that it points outward of the channel domain $\Omega$. 
% This velocity profile is of course an idealization, and its implications on our results will be examined further. 
%Although a parabolic profile may not accurately represent the physically realistic inlet conditions in micro- and mini-channel applications, this idealized condition allows to assess and compare the validity of the developed macro-scale description for various Reynolds numbers and offset strip fin channel geometries. 
At the outlet $\Gamma_{\text{out}}$, the pressure is set to a constant reference pressure equal to zero. 
Additionally, a no-slip boundary condition is imposed at the fluid-solid interface $\Gamma_{fs}$, the top boundary $\Gamma_{t}$, the bottom boundary $\Gamma_{b}$ and the side-wall boundaries $\Gamma_{\text{sides}}$. \\

The macro-scale velocity field $\langle \boldsymbol{u} \rangle_{m}$ and macro-scale pressure field $\langle p \rangle_{m}$ are obtained by applying a double volume-averaging operator $\langle \; \rangle_{m}$ on the velocity distribution $\boldsymbol{u}$ and pressure distribution $p$. 
The latter operator expresses a convolution product in $\mathbb{R}^{3}$ with a weighting function $m$ such that $\langle \phi \rangle_{m} = m \ast \phi$ and (Refs.~\onlinecite{quintard1994transport2,buckinx2015multi}) 
\begin{equation}
  m \left( \boldsymbol{y} \right) = \frac{1}{l_{3}} \text{rect} \left( \frac{y_{3}}{l_{3}} \right) \prod_{j=1}^{2} \frac{l_{j} - 2 | y_{j} | }{l_{j}} \text{rect} \left( \frac{y_{j}}{2 l_{j}} \right).
\label{eq:weightingfunction}
\end{equation}
Here, $\text{rect}$ denotes a normalized rectangle function as defined in (Refs.~\onlinecite{weisstein2002rectangle}). 
This filter operator has a filter window equal to the local unit cell $\Omega_{\text{unit}}^{2 \times 2} \left( \boldsymbol{x} \right)$, as defined in (Ref.~\onlinecite{buckinx2022arxiv}) and illustrated in Figure \ref{fig:geometrydns}. 
% defined as 
% \begin{equation}
%   \Omega_{\text{unit}}^{n_{1} \times n_{2}} \left( \boldsymbol{x} \right) \triangleq \Biggl\{ \boldsymbol{r} = \boldsymbol{x} + \boldsymbol{y} | \exists c_{j} \in \left[ -\frac{1}{2} \text{,} \frac{1}{2} \right] \Leftrightarrow \boldsymbol{y} = \sum_{j=1}^{2} c_{j} n_{j} \boldsymbol{l}_{j} + c_{3} \boldsymbol{l}_{3} \Biggl\}. 
% \label{eq:filterwindow}
% \end{equation}
When the flow in the channel is steady, the macro-scale flow variables satisfy the following macro-scale flow equations (Refs.~\onlinecite{quintard1994transport2,buckinx2017macro,buckinx2022arxiv}):
\begin{equation}
\begin{aligned}
  \nabla \cdot \langle \boldsymbol{u} \rangle_{m} &= 0,  \\
  \rho_{f} \nabla \cdot \left( \epsilon_{fm}^{-1} \langle \boldsymbol{u} \rangle_{m} \langle \boldsymbol{u} \rangle_{m} \right) &= - \nabla \langle p \rangle_{m} + \mu_{f} \nabla^2 \langle \boldsymbol{u} \rangle_{m} + \boldsymbol{f}_{\text{closure}} , \\
\end{aligned}
\label{eq:NavierStokesMS}
\end{equation}
where, the total macro-scale closure force $\boldsymbol{f}_{\text{closure}}$ can be written as 
\begin{equation}
    \boldsymbol{f}_{\text{closure}} = -\rho_{f} \nabla \cdot \boldsymbol{M} + \boldsymbol{b}. 
    \label{eq:totalclosureforce}
\end{equation}
The first contribution in (\ref{eq:totalclosureforce}) originates from the macro-scale momentum dispersion tensor $\boldsymbol{M} \triangleq \langle \boldsymbol{u} \boldsymbol{u} \rangle_{m} - \epsilon_{fm}^{-1} \langle \boldsymbol{u} \rangle_{m} \langle \boldsymbol{u} \rangle_{m}$. 
Here, the weighted porosity is defined as $\epsilon_{fm} \triangleq \langle \gamma_{f} \rangle_{m}$, while the fluid indicator function is defined by $\gamma_{f}(\boldsymbol{x}) = 1 \leftrightarrow \boldsymbol{x} \in \Omega_{f}$, $\gamma_{f}(\boldsymbol{x}) = 0 \leftrightarrow \boldsymbol{x} \in \Omega_{s}$. 
The second contribution is the macro-scale closure force, 
\begin{equation}
    \boldsymbol{b} \triangleq \langle \boldsymbol{n}_{fs} \cdot \left( -p_{f} \boldsymbol{I} + \boldsymbol{\tau}_{f} \right) \delta_{fs} \rangle_{m}, 
    \label{eq:closureforce}
\end{equation}
which results from the no-slip boundary condition on the fluid-solid interface $\Gamma_{fs}$.
%%Here, $\boldsymbol{I}$ is the identity tensor and $\boldsymbol{\tau}_{f} = \mu_{f} \left( \nabla^{\nu} \boldsymbol{u} + \nabla^{\nu} \boldsymbol{u}^{T} \right)$ the viscous stress tensor with $\nabla^{\nu}$ the gradient operator in the usual sense (Refs.~\onlinecite{schwartz1966theorie,gagnon1970distribution}).
Here, $\boldsymbol{I}$ is the identity tensor and $\boldsymbol{\tau}_{f} = \mu_{f} \left( \nabla \boldsymbol{u}_{f} + \nabla \boldsymbol{u}_{f}^{T} \right)$ the viscous stress tensor. 
Further, $\boldsymbol{n}_{fs}$ denotes the normal vector pointing from $\Omega_{f}$ to $\Omega_{s}$ and $\delta_{fs}$ is the Dirac surface indicator of $\Gamma_{fs}$. 
We remark that, in order to determine $\langle \boldsymbol{u} \rangle_{m}$ and $\langle p \rangle_{m}$ on the entire channel domain $\Omega$, the variables $\boldsymbol{u}$ and $p$ are defined as extended distributions based on $\boldsymbol{u}_{f}$ and $p_{f}$, respectively. 
More specifically, the extended velocity field $\boldsymbol{u}$ is defined as: $\boldsymbol{u} = \boldsymbol{u}_{f}$ in $\Omega_{f}$, $\boldsymbol{u} = 0$ in $\Omega_{s}$ and $\boldsymbol{u} = \boldsymbol{u}_{e}$ in $\mathbb{R}^{3} \setminus \Omega$. 
Similarly, the extended pressure field $p$ is defined as: $p = p_{f}$ in $\Omega_{f}$, $p = 0$ in $\Omega_{s}$ and $p = p_{e}$ in $\mathbb{R}^{3} \setminus \Omega$. 
The extensions $\boldsymbol{u}_{e}$ and $p_{e}$ in this work are chosen identical to those in (Ref.~\onlinecite{buckinx2022arxiv}), in order to ensure the validity of the macro-scale flow equations (\ref{eq:NavierStokesMS}).

\subsection{Numerical procedure}

% The same discretization schemes and algorithms as for the periodically developed flow were used to solve the governing equations. (Ref.~\onlinecite{vangeffelen2021friction})
The channel flow equations (\ref{eq:NavierStokes})-(\ref{eq:BoundaryConditions}) have been solved numerically by means of the software package FEniCSLab developed by G. Buckinx in the finite-elements-based open-source computing platform FEniCS (Ref.~\onlinecite{AlnaesBlechta2015a}). 
In this package, the parallel fractional-step solver of \textit{Oasis} for the unsteady incompressible Navier-Stokes equations, developed by Mortensen and Valen-Sendstad (Ref.~\onlinecite{mortensen2015oasis}), was re-implemented using an object-oriented approach. 
The offset strip fin channel domain has been spatially discretized on a structured mesh containing between 26,530,560 and 193,294,080 grid cells, depending on the geometrical parameters of the channel. 
% In each local unit cell domain $\Omega_{\text{unit}}$, this corresponds to a number of grid cells equal to 48 along $\boldsymbol{e}_{1}$, 26 to 108 along $\boldsymbol{e}_{2}$, and 13 to 104 along $\boldsymbol{e}_{3}$. 
On this mesh, the velocity and pressure fields have been discretized by continuous Galerkin tetrahedral elements of the second and first order, respectively. 
To obtain the numerical flow solution for a single geometry and Reynolds number, a computational time of up to 12 hours was required on 10 nodes of each 36 processors (Xeon Gold 6140 2.3GHz with 192GB of RAM), such that after a total number of 2000 discrete time steps from a zero initial velocity field, a steady state was observed. % at time $t' = 20 l / u_{b}$
The time-stepping stability was ensured by constricting the local Courant–Friedrichs–Lewy number in each mesh cell to a value below 0.9. 
According to our mesh-independence study, the discretization errors on the velocity profiles are below 3\% for all the cases presented in this work. 
For the double volume-averaging operations, an explicit finite-element integral operator, implemented in FEniCSLab by G. Buckinx (Ref.~\onlinecite{buckinx2022arxiv}), has been employed. 
Approximately 6 hours on 10 nodes of 36 processors (Xeon Gold 6140 2.3GHz with 192GB of RAM) were needed to filter each scalar flow variable with the double volume-averaging operator. 

\begin{figure}[ht]
\includegraphics[scale = 1.25]{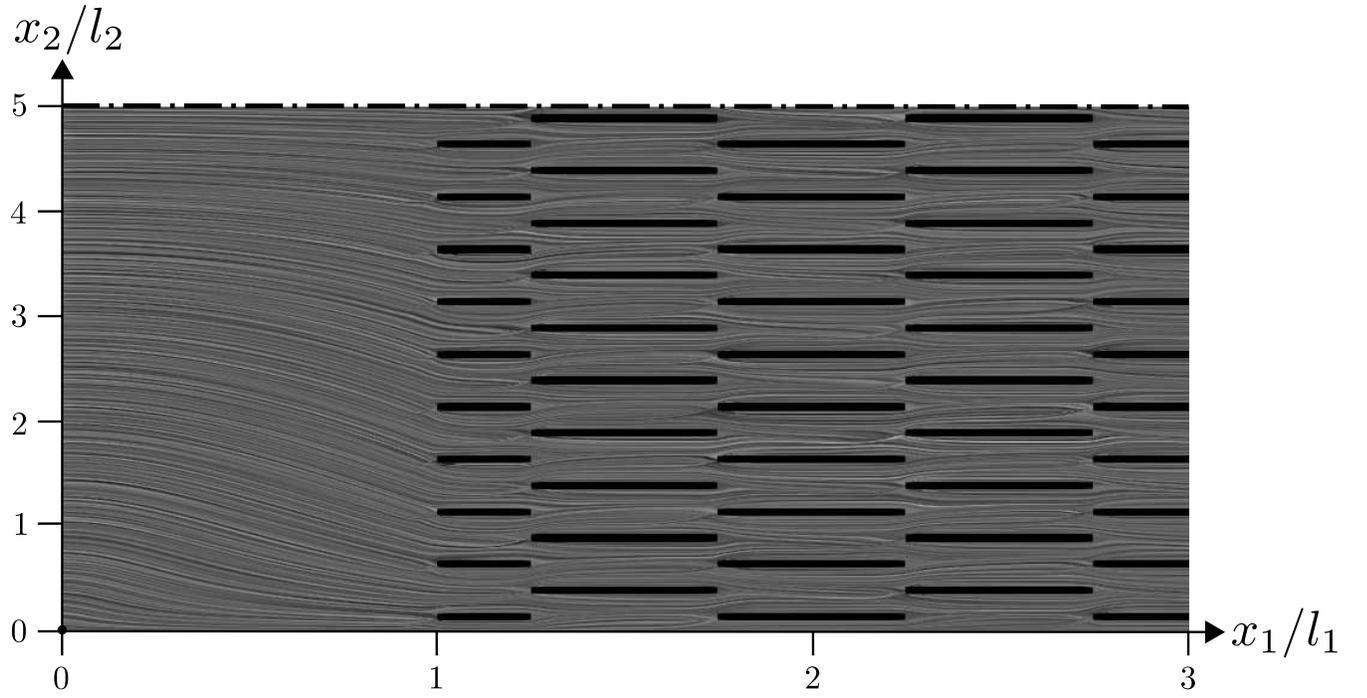}
\caption{\label{fig:streamlinesRE600T4H12S24N10} Example of flow pattern visualization by Line Integral Convolution (Ref.~\onlinecite{loring2014screen}) in half of the cross-section mid-plane spanned by $\boldsymbol{e}_{1}$ and $\boldsymbol{e}_{2}$, when $Re_{b}=192$, $h/l=0.12$, $s/l=0.24$, $t/l=0.04$, $s_0=l_1$, $s_N=2.5 l_1$, $N_{1}=20$, and $N_{2}=10$}
\end{figure}

In Figure \ref{fig:streamlinesRE600T4H12S24N10}, the simulated velocity field in an offset strip fin channel is shown as an example, in this case for a Reynolds number $Re_{b} \triangleq \rho_{f} u_{b} (2 L_{3} ) / \mu_{f} = \rho_{f} u_{b} 2 (h + t) / \mu_{f} = 192$ while $h/l=0.12$, $s/l=0.24$, $t/l=0.04$, $s_0=l_1$, $s_N=2.5 l_1$, $N_{1}=20$, and $N_{2}=10$. 
% $Re_{b} \triangleq \rho_{f} u_{b} (2 h) / \mu_{f} = 144$
In this figure, the detailed flow patterns in the mid-plane of the channel, spanned by $\boldsymbol{e}_{1}$ and $\boldsymbol{e}_{2}$, are visualized by means of Line Integral Convolution (Ref.~\onlinecite{loring2014screen}). 
Only half the region near the inlet of the channel is illustrated for the sake of clarity, although the array geometry and thus the velocity field is not symmetric with respect to the plane $x_2=L_2/2$. 
It can be seen that, in this fin array, the flow patterns around the offset strip fins become qualitatively similar after a short distance (at $x_1 \simeq 2 l_1$) from the start of the fin array. \\
% , in contrast with arrays of a less porous fin geometry such as square cylinders (Ref.~\onlinecite{buckinx2022arxiv})

% color range 0-5
% Number of steps: 100
% Step size: 0.5
% Enhance contrast: color only
% LIC intensity: 1

\newpage
\clearpage

\section{\label{sec:onsetMS}Onset of developed macro-scale flow}

According to the macro-scale description of Buckinx and Baelmans (Ref.~\onlinecite{buckinx2015multi}), which is based on the filter (\ref{eq:weightingfunction}), developed macro-scale flow is established once the flow has become periodically developed in the main flow direction $\boldsymbol{e}_{1}$. 
As a consequence, the onset of developed macro-scale flow in channels with offset strip fin arrays can be specified by the onset of streamwise periodically developed flow in these channels.

\subsection{\label{sec:onsetperiodic}Onset of streamwise periodically developed flow}

Based on the flow fields obtained via DNS, we have determined the onset of streamwise periodically developed flow for a wide range of Reynolds numbers and geometrical parameters of the channel. 
The onset point $x_{\text{periodic}}$ is computed as the coordinate $x_{1}$ after which the velocity field agrees with the streamwise periodically developed solution within an accuracy of 1\%, such that for $x_{1} \geqslant x_{\text{periodic}}$: $\boldsymbol{u} \left( \boldsymbol{x} + \boldsymbol{l}_{1} \right) \simeq \boldsymbol{u} \left( \boldsymbol{x} \right)$. 

\begin{figure}[ht!]
\centering
\begin{minipage}{.475\textwidth}
% \centering
\raggedleft
\includegraphics[scale = 1.00]{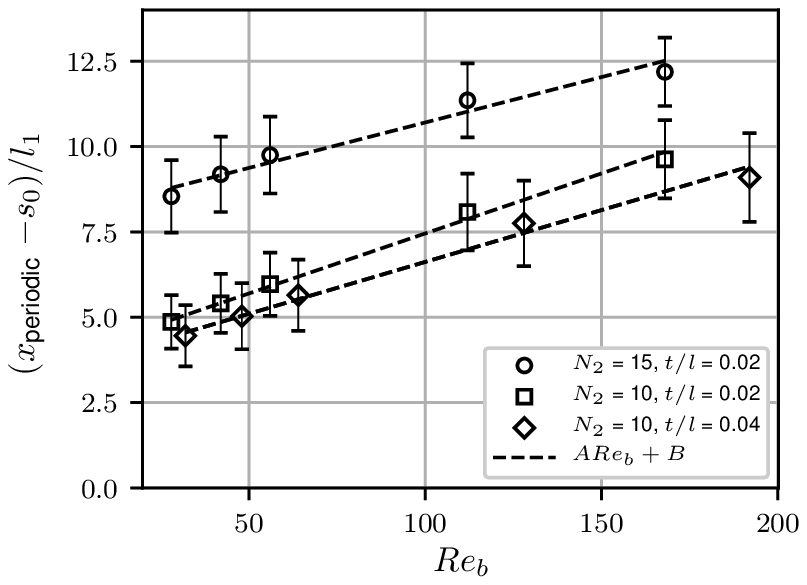}
\caption{\label{fig:xdev_RE_t} Influence of the Reynolds number on the onset of streamwise periodically developed flow, when $h/l=0.12$, $s/l=0.48$}
% epsilon = 0.692 - 0.823
\end{minipage}
\hfill
\begin{minipage}{.475\textwidth}
% \centering
\raggedright
% \raggedleft
\includegraphics[scale = 1.00]{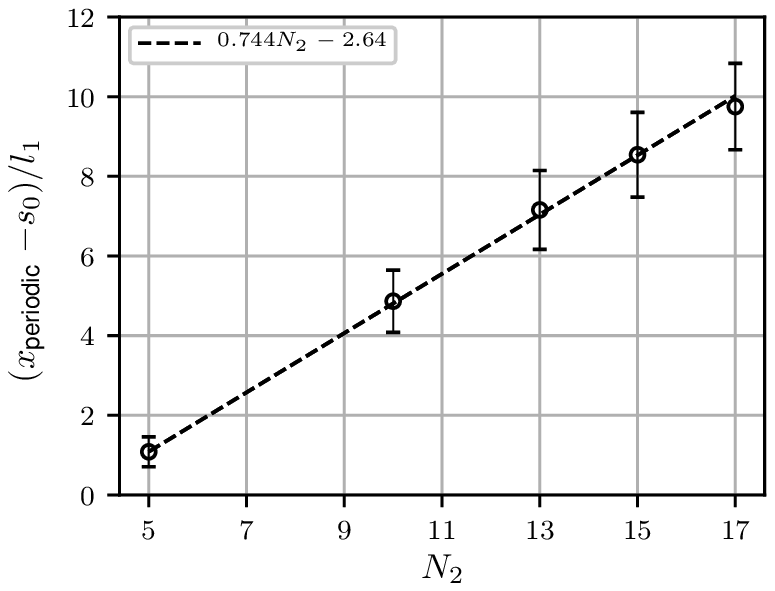}
\caption{\label{fig:xdev_N2} Influence of the channel aspect ratio on the onset of streamwise periodically developed flow, when $Re_{b}=28$, $h/l=0.12$, $s/l=0.48$, $t/l=0.02$}
% epsilon = 0.823
\end{minipage}
\end{figure}

Figure \ref{fig:xdev_RE_t} shows the dependence of the onset point $x_{\text{periodic}}$ on the Reynolds number $Re_{b}$ for two different fin thickness-to-length ratios $t/l$ and two different channel aspect ratios $L_{3}/L_{2}=l_{3}/(N_{2} l_{2})$. 
In this figure, also the numerical uncertainty due to discretization errors is indicated by means of error bars. % 99\%-level
% These uncertainty bars are calculated by perturbating the accuracy constraint of 1\%. 
All the shown data points are captured by a linear relationship with a maximum error below 3\%. 
We have $(x_{\text{periodic}} - s_{0}) / l_{1} \simeq 0.0266 Re_{b} + 8.04$ when $\{ N_{2}=15, t/l=0.02 \}$, $(x_{\text{periodic}} - s_{0}) / l_{1} \simeq 0.0351 Re_{b} + 3.94$ when $\{ N_{2}=10, t/l=0.02 \}$, and $(x_{\text{periodic}} - s_{0}) / l_{1} \simeq 0.0304 Re_{b} + 3.58$ when $\{ N_{2}=10, t/l=0.04 \}$. 
% We have $(x_{\text{periodic}} - s_{0}) / l_{1} \simeq 0.00745 Re_{b} + 8.04$ when $\{ N_{2}=15, t/l=0.02 \}$, $(x_{\text{periodic}} - s_{0}) / l_{1} \simeq 0.00984 Re_{b} + 3.94$ when $\{ N_{2}=10, t/l=0.02 \}$, and $(x_{\text{periodic}} - s_{0}) / l_{1} \simeq 0.00972 Re_{b} + 3.58$ when $\{ N_{2}=10, t/l=0.04 \}$. 
A similar linear scaling of the dimensionless flow development length $x_{\text{periodic}}/ l_{1}$ with the Reynolds number has been observed for flows in micro- and mini-channels without solid structures, though at Reynolds numbers above fifty (Refs.~\onlinecite{atkinson1969low,chen1973flow,schlichting1997boundary,sadri2002accurate,durst2005development,ahmad2010experimental}). 
% For smaller Reynolds numbers, however, the dependence becomes nonlinear due to the transition between dominant eigenvalues that characterize the flow development, as reported in the work of Sadri and Floryan (Ref.~\onlinecite{sadri2002accurate}). 
Furthermore, a linear scaling between $x_{\text{periodic}}/ l_{1}$ and $Re_{b}$ over the same range $Re_{b} \in \left( 50,300 \right)$ has also been reported for micro- and mini-channels containing arrays of square cylinders in the study of Buckinx (Ref.~\onlinecite{buckinx2022arxiv}). 

% Physical explanation: 
The linear form of the previous correlations, $x_{\text{periodic}} / l_{1} \simeq A Re_{b} + B$, indicates that the onset point $x_{\text{periodic}}$ becomes asymptotically independent of the bulk velocity in the limit of $Re_{b} \rightarrow 0$, i.e. when viscous diffusion becomes dominant over advection (or flow inertia). 
This can be understood from the notion that, in that limit, the flow development will mainly  occur through viscous diffusion along the main flow direction.
As such, the development will take place over a finite length scale which is solely determined by the rate of viscous diffusion, and therefore independent of the bulk velocity 
(Refs.~\onlinecite{vrentas1966effect,atkinson1969low,durst2005development}).
On the other hand, for larger Reynolds numbers, the linear form indicates that the onset point will increase almost proportionally to the Reynolds number: $x_{\text{periodic}}/ l_{1} \sim A Re_{b}$. 
In that case, the flow development is primarily driven by viscous diffusion in the directions perpendicular to the main flow. 
As such, the flow development takes place over a length scale which is determined by the ratio of transversal diffusion to streamwise advection by the bulk velocity, as expressed by $Re_{b}$.
We remark that this asymptotic scaling law $x_{\text{periodic}}/ l_{1} \simeq A Re_{b}$ for higher Reynolds numbers is also observed for flow development in external boundary-layer flows (Ref.~\onlinecite{schlichting1997boundary}).
%which also occur due from a balance involve diffusive transport perpendicular to the streamwise direction. % and convective transport along the streamwise direction. 
%%We remark that for the transitional flow regime at $Re_{b} \gg 1000$, the flow development length, corresponding to the onset point to vortex shedding, will be mainly determined by the convective transport and will decrease for an increasing value of the Reynolds number (Refs.~\onlinecite{joshi1987heat,mochizuki1988flow,dejong1997experimental}). 
%%For both channels with and without fins, the linear scaling can be explained from the eigenvalues of the quasi-developed flow modes in the entrance region of the channel (Refs.~\onlinecite{sadri2002accurate,buckinx2022arxiv}). 
%%These modes and their eigenvalues will be discussed in detail in Section \ref{sec:onsetquasi}. 

Figure \ref{fig:xdev_RE_t} further suggests that the dependence of $x_{\text{periodic}}$ on the thickness-to-length ratio $t/l$ of the offset strip fin is not significant for micro- and mini-channel applications. 
However, it can be expected that the onset of streamwise periodically developed flow will move upstream if $t/l$ increases, as can still be recognized from the data in this figure. 
This can be explained by the fact that for larger values of $t/l$, the porosity decreases and the fins block a larger cross-sectional area in the flow. 
Due to this blockage effect, the diffusive transport perpendicular to the main flow direction becomes more significant than it is in fin arrays of a higher porosity. 

In Figure \ref{fig:xdev_N2}, it can be observed that the onset point $x_{\text{periodic}}$ becomes larger as the number of unit cells in the lateral direction $N_{2}$ increases, and thus the channel aspect ratio decreases. 
More specifically, the relation between $x_{\text{periodic}}$ and the aspect ratio in the former figure can be predicted by a linear fit within a relative error of 3\%. 
Similar linear correlations between the flow development length and aspect ratio have been proposed for rectangular micro- and mini-channels without fins (Refs.~\onlinecite{li2019flow,ferreira2021hydrodynamic}), when one converts the definitions of the flow development length and Reynolds number which they rely on, to those used in this work. 
The linear relationship between $x_{\text{periodic}}$ and $N_{2}$ can be attributed to the fact that the length scale for lateral viscous diffusion is proportional to $N_{2} l_{2}$ so that it takes a longer distance for the flow to develop when $N_{2}$ is larger.
%so that any perturbation of the velocity profile at the onset of the offset strip fin array with respect to the developed velocity field will need to be transported by diffusion along the lateral direction over a length proportional to $N_{2}$. %, i.e. from the channel side wall to the channel centerline. 
%When the unit cell geometry and porosity remain the same, the lateral diffusive transport also remains constant. 
%Consequently, when $N_{2}$ increases, the development length in the offset strip fin channel increases. 

\begin{figure}[ht!]
\centering
\begin{minipage}{.475\textwidth}
% \centering
\raggedleft
\includegraphics[scale = 1.00]{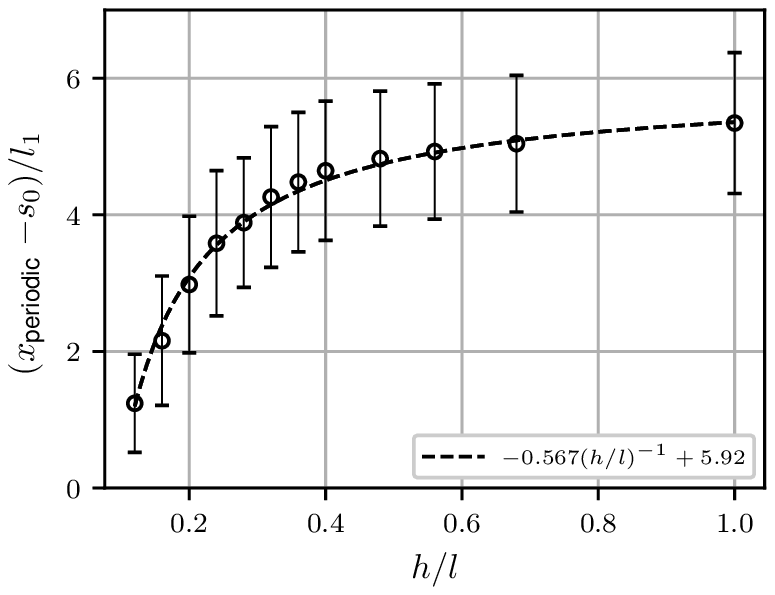}
\caption{\label{fig:xdev_h} Influence of the fin height-to-length ratio on the onset of streamwise periodically developed flow, when $Re_{b} l / (2 L_{3}) = \rho_{f} u_{b} l / \mu_{f} = 600$, $N_{2}=10$, $s/l=0.12$, $t/l=0.02$}
% epsilon = 0.735 - 0.840
\end{minipage}
\hfill
\begin{minipage}{.475\textwidth}
% \centering
\raggedright
% \raggedleft
\includegraphics[scale = 1.00]{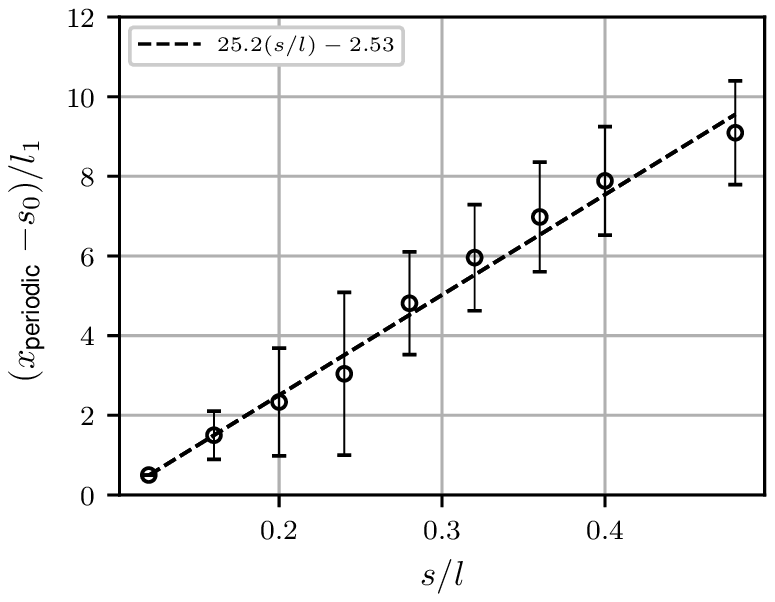}
\caption{\label{fig:xdev_s} Influence of the fin pitch-to-length ratio on the onset of streamwise periodically developed flow, when $Re_{b}=192$, $N_{2}=10$, $h/l=0.12$, $t/l=0.04$}
% epsilon = 0.563 - 0.692
\end{minipage}
\end{figure}

On the other hand, the relationship between $x_{\text{periodic}}$ and the dimensionless fin height $h/l$ (or fin height-to-length ratio) approximately satisfies an inversely linear form, as it can be seen in Figure \ref{fig:xdev_h}. 
At least, the form $ x_{\text{periodic}}/l_1 \simeq A (h/l)^{-1} + B$ holds when the bulk velocity $u_{b}$ and fin length $l$ are kept constant, as we have chosen $Re_{b} l / (2 L_{3}) = \rho_{f} u_{b} l / \mu_{f} = 600$ in Figure \ref{fig:xdev_h}, which implies that $Re_{b} \in (168,1224)$ for $h/l \in (0.12,1)$. 
The inversely linear form results in a maximum relative error of 4\% for the onset point data in the former figure. 
It clearly reflects that the onset point moves upstream in the channel when $h/l$ decreases.
This can be attributed to the fact that for relatively lower fin heights, the porosity decreases, so that lateral viscous diffusion becomes again more significant due to flow blockage along the main flow direction.
On top of that, the form indicates that the onset point becomes independent of $h/l$ once the fin height-to-length ratio is sufficiently large ($h/l > 0.8$).
In that case, the flow has become rather two-dimensional (Ref.~\onlinecite{vangeffelen2021friction}), such that the development length is no longer affected by the presence of the top and bottom walls, nor the distance $h/l$  between them. 
%This asymptotic trend for $h/l \rightarrow \infty$ is also captured by the proposed correlation form. 
% This form reflects that the onset point becomes independent of $h/l$ when $h/l$ increases, and results in a maximum relative error of 4\% with the onset point data in this figure. 

As we have illustrated in Figure \ref{fig:xdev_s}, also the scaling of $x_{\text{periodic}}$ with the fin pitch-to-length ratio $s/l$ is found to be linear, within a numerical uncertainty of 10\%. 
Furthermore, when the dimensionless fin pitch $s/l$ increases, the onset point of streamwise periodically developed flow moves more downstream.
The fin pitch-to-length ratio $s/l$ thus has a similar influence on the flow development as the number of unit cells in the lateral direction $N_{2}$.
This observation may not come as a surprise, since both affect the length scale for lateral viscous diffusion in the same way as described before.
Besides, when the ratio $s/l$ increases, also the porosity of the array increases, so that advection in the main flow direction outbalances lateral viscous diffusion.

% Given that the total influence of the relative fin pitch can be captured by a linear correlation form, it can be expected that the latter influence is inferior to the influence of $s/l$ on the channel aspect ratio. 
% This linear dependence on $s/l$ is thus similar to the linear dependence on $N_{2}$. 
% The observed relationships between $x_{\text{periodic}}$ and the geometrical parameters $N_{2}$, $h/l$ and $s/l$ 
% % correspond to an inverse linear dependence of $x_{\text{periodic}}$ with the channel's aspect ratio $L_{3}/L_{2} \simeq (h/l)/ \left( 2 N_{2} (s/l) \right)$. This is
% are all in agreement with the correlation proposed for rectangular micro- and mini-channels without fins in the work (Ref.~\onlinecite{ferreira2021hydrodynamic}), taking into account the conversion of their definitions of the development length and Reynolds number to those used in this work. 

As a consequence of the previous observations, it can be generally stated that the onset of streamwise periodically developed flow moves further downstream as the porosity $\epsilon_{f}$ of the fin array increases, for all flow conditions and geometrical parameters considered in this work. 
In addition, the previously observed trends characterizing the influence of the geometrical parameters are in agreement with those for arrays of square cylinders (Ref.~\onlinecite{buckinx2022arxiv}). 
In particular, for arrays of square cylinders, the influence of the channel aspect ratio can also be accurately captured by a linear relationship between $x_{\text{periodic}}$ and $N_{2}$. 
Moreover, in (Ref.~\onlinecite{buckinx2022arxiv}), it is shown that the onset of periodically developed flow moves downstream when the channel height $h/l_{1}$ increases, and becomes independent of $h/l_{1}$ for sufficiently large channel heights ($h/l_{1} > 1$). 
Overall, it can be concluded that the flow development length $(x_{\text{periodic}} - s_{0})$ for offset strip fin arrays in micro- and mini-channels remains rather small in comparison to fin arrays of a higher porosity, like arrays of square cylinders (Ref.~\onlinecite{buckinx2022arxiv}), when we consider similar ranges of the channel height, channel aspect ratio and Reynolds number. 
After all, the flow development lengths from this work range from 1 to 12 unit cell lengths for an array porosity between 0.56 and 0.84, while those from (Ref.~\onlinecite{buckinx2022arxiv}) range from 10 to 60 unit cell lengths for an array porosity between 0.75 and 0.94. 
% OSF:    $\epsilon_{f} \simeq 0.56 - 0.84$, $(x_{\text{periodic}} - s_{0})/l_{1} \simeq 1  - 12$
% square: $\epsilon_{f} \simeq 0.75 - 0.94$, $(x_{\text{periodic}} - s_{0})/l_{1} \simeq 10 - 60$

As we will show in Section \ref{sec:onsetquasi}, the previously observed trends for the onset point of periodically developed flow can be explained mainly by the scaling laws for the eigenvalues of the quasi-developed flow in the entrance region of the offset strip fin channel.


\newpage
\clearpage

Next to the onset point, we have also determined the end point of streamwise periodically developed flow in the offset strip fin array.
This end point $x_{\text{end}}$ has been determined as the coordinate $x_{1}$ after which the relative difference between the local velocity and the streamwise periodically developed solution again becomes larger than 1\%. 
For all the cases considered in this study, the end point practically coincides with the end of the offset strip fin array such that $x_{\text{end}} \simeq L_{1} - s_{N}$. 
This suggests that the streamwise periodicity of the flow is not significantly affected by the outlet region (and outlet boundary conditions) for the micro- and mini-channels considered here. 
% We remark that, even for different values of $s_{N}$, the streamwise periodicity was preserved until the final fin row for the considered range of Reynolds numbers and geometrical parameters. 
The same observation has been made for channels with an array of square cylinders in the work of Buckinx (Ref.~\onlinecite{buckinx2022arxiv}).

\subsection{\label{sec:onsetdevmacro} Region of developed macro-scale flow}

\begin{figure}[ht]
\includegraphics[scale = 0.95]{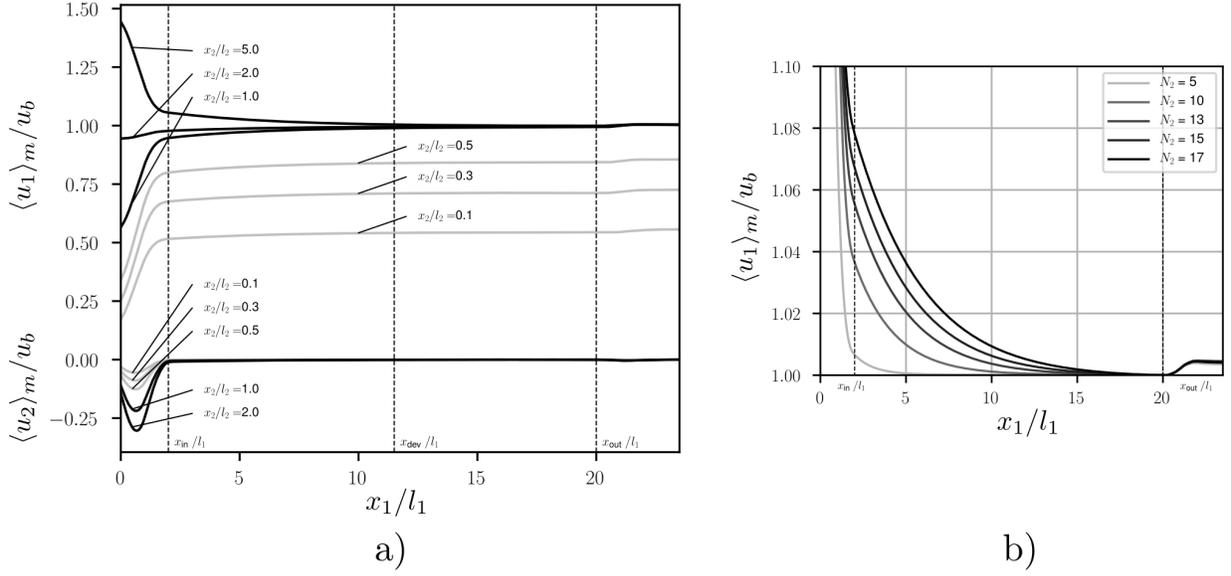}
\caption{\label{fig:msvel} Streamwise and lateral macro-scale velocity components outside (black) and inside (grey) the side-wall region, when $Re_{b}=168$, $h/l=0.12$, $s/l=0.48$, $t/l=0.02$, $s_{0}=l_{1}$, $s_{N}=2.5 l_{1}$, $N_{1}=20$, and $N_{2}=10$ (a), and streamwise macro-scale velocity component along the channel centerline ($x_{2} = L_{2}/2$), when $Re_{b}=28$, $h/l=0.12$, $s/l=0.48$, $t/l=0.02$, $s_{0}=l_{1}$, $s_{N}=2.5 l_{1}$, and $N_{1}=20$ (b)}
\end{figure}

Strictly speaking, the macro-scale velocity becomes independent of the coordinate $x_{1}$, and thus truly developed, only at a distance $l_{1}$ after the onset of streamwise periodically developed flow, since $l_{1}$ is the radius of the chosen filter window $\Omega_{\text{unit}}^{2 \times 2}$.
So, the region of developed macro-scale flow $\Omega_{\text{dev}}$, in the strict sense, is given by $x_{1} \in \left( x_{\text{dev}}, x_{\text{end}} - l_{1} \right)$, with $x_{\text{dev}} = x_{\text{periodic}} + l_{1}$. 
The extent of this region $\Omega_{\text{dev}}$ is illustrated in Figure \ref{fig:msvel}(a), together with the streamwise profiles of the macro-scale velocity components $ \langle u_{1} \rangle_{m}$ and $\langle u_{2} \rangle_{m}$. 
As the former figure shows, the macro-scale velocity profile can be written as $\langle \boldsymbol{u} \rangle_{m} \left( \boldsymbol{x} \right) = U_{\text{dev}} \left( x_{2} \right) \boldsymbol{e}_{1}$ in the region $\Omega_{\text{dev}}$. 
From a practical point of view, the approximation $\langle \boldsymbol{u} \rangle_{m} \left( \boldsymbol{x} \right) \simeq U_{\text{dev}} \left( x_{2} \right) \boldsymbol{e}_{1}$ actually holds quite well over the entire fin array, except the first and last fin rows, according to our DNS results.

For all the Reynolds numbers and geometries considered in this work, we observe that the largest deviations from the developed profile $\| \langle \boldsymbol{u} \rangle_{m} - U_{\text{dev}} \| / \| U_{\text{dev}} \|$ are below 8\% if we do not consider the first and last rows of the array, where streamwise porosity gradients occur. 
As such, the region of developed macro-scale flow practically coincides with the region where the weighted porosity $\epsilon_{fm}$ of the fin array does not vary with the streamwise coordinate $x_{1}$, which corresponds to $x_{1} \in (x_{\text{in}},x_{\text{out}})$ with $x_{\text{in}} = s_{0} + l_{1}$ and $x_{\text{out}} = L_{1} - \left( s_{N} + l_{1} \right)$. 
This is especially true for larger values of the channel aspect ratio $L_{3}/L_{2}$, as it can be seen in Figure \ref{fig:msvel}(b). 
In particular, for all our data, the deviations $\| \langle \boldsymbol{u} \rangle_{m} - U_{\text{dev}} \| / \| U_{\text{dev}} \|$ remain below 5\% for $x_{1} \in (x_{\text{in}},x_{\text{out}})$ when $N_{2} \leqslant 10$. 
% the macro-scale velocity is only a function of the $x_{2}$ coordinate in the channel for : $\langle \boldsymbol{u} \rangle_{m} \left( \boldsymbol{x} \right) = U_{\text{dev}} \left( x_{2} \right) \boldsymbol{e}_{1}$. 
The small deviations are also reflected by the fact that the transversal component of the macro-scale velocity component $\langle u_{2} \rangle_{m}$ remains virtually zero throughout the entire offset strip fin array, as visible in Figure \ref{fig:msvel}(a). 
Consequently, the flow angle of attack $\alpha \triangleq \arctan \left( \langle u_{2} \rangle_{m} / \langle u_{1} \rangle_{m} \right)$ does not exceed 1$^{\circ}$ for $x_{1} \in (x_{\text{in}},x_{\text{out}})$. 
The same conclusion can be drawn for all the other flow conditions and channel geometries examined in this study. 

%%%new paragraph
Despite the fact that the macro-scale flow in the fin array can be treated as fully developed from a practical perspective, flow development is still clearly distinguishable at the macro-scale level, especially for small channel aspect ratios. %, according to Figure \ref{fig:msvel}. 
In Figure \ref{fig:msvel}, the region of developing macro-scale flow $\Omega_{\text{predev}}$ has been identified by $x_{1} \in (x_{\text{in}},x_{\text{dev}})$, as we exclude the inlet region $(0,x_{\text{in}})$ before the fin array. 
% developing flow effects at the macro-scale can be ignored, is also reflected by the fact that the transversal macro-scale velocity component remains virtually zero in Omega_predev, and the fact that the angle of attack remains small
% Similar to the developed region, the macro-scale velocity component $\langle u_{2} \rangle_{m}$ also remains close to zero in the developing region, resulting in a flow angle of attack $\alpha \triangleq \arctan \left( \langle u_{2} \rangle_{m} / \langle u_{1} \rangle_{m} \right)$ not exceeding 1$^{\circ}$ throughout the entire offset strip fin array. 
% The same conclusion can be drawn for all the other offset strip fin channels examined in this study. 
% Moreover, the macro-scale velocity component $\langle u_{1} \rangle_{m}$ does not vary significantly in $\Omega_{\text{predev}}$, which is in agreement with the small flow development lengths for offset strip fin micro- and mini-channels computed in this work. 

% \begin{figure}[ht]
% \includegraphics[scale = 1.00]{figures/MSvelocityRE600T2H12S48N10.eps}
% \caption{\label{fig:msvel} Streamwise and lateral macro-scale velocity components outside (black) and inside (grey) the side-wall region, when $Re_{b}=600$, $h/l=0.12$, $s/l=0.48$, $t/l=0.02$, $s_{0}=l_{1}$, $s_{N}=2.5 l_{1}$, $N_{1}=20$, and $N_{2}=10$}
% \end{figure}

% \begin{figure}[ht]
% \includegraphics[scale = 1.00]{figures/u_filt_N2_MS.eps}
% \caption{\label{fig:u_filt_N2_MS} Streamwise macro-scale velocity component along the channel centerline ($x_{2} = L_{2}/2$), when $Re_{b}=100$, $h/l=0.12$, $s/l=0.48$, $t/l=0.02$, $s_{0}=l_{1}$, $s_{N}=2.5 l_{1}$, and $N_{1}=20$}
% \end{figure}

\newpage
\clearpage

\subsection{\label{sec:accuracy}Accuracy of the developed friction factor correlation}

\begin{figure}[ht]
\includegraphics[scale = 1.00]{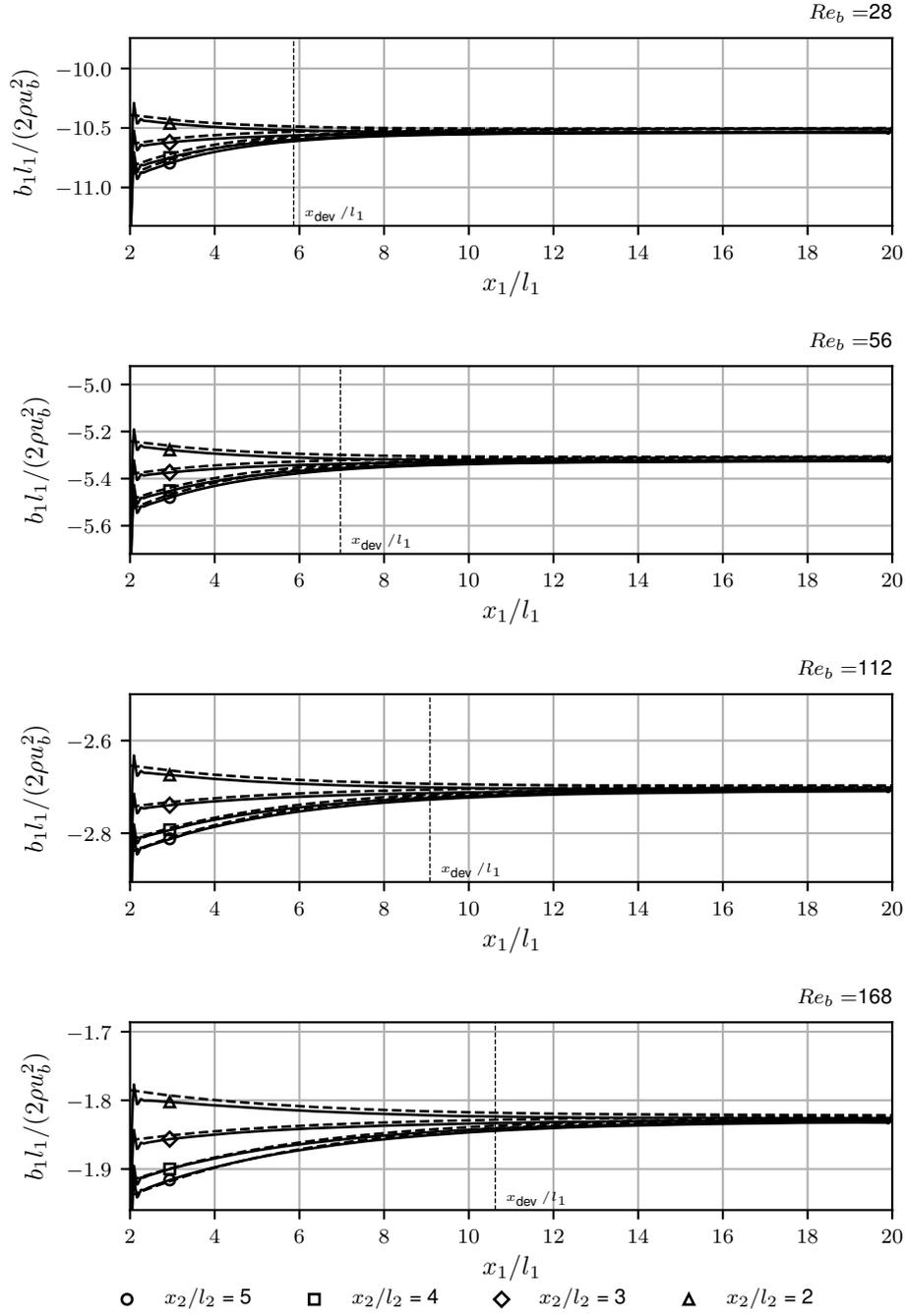}
\caption{\label{fig:b_x_t0.02_h0.12_s0.48} Macro-scale closure force $\boldsymbol{b}$ (full) and its prediction $\boldsymbol{b}_{\text{unit}}$ (dashed) by the developed correlation from (Ref.~\onlinecite{vangeffelen2021friction}), when $h/l=0.12$, $s/l=0.48$, $t/l=0.02$, $s_{0}=l_{1}$, $s_{N}=2.5 l_{1}$, $N_{1}=20$, and $N_{2}=10$}
\end{figure}

%% intro
%In this section, the discrepancies between the macro-scale closure force in offset strip fin micro- and mini-channels and the developed friction factor correlation from (Ref.~\onlinecite{vangeffelen2021friction}) is assessed. 
%This is done in both the developed and developing macro-scale flow region. 
In order to assess the accuracy of the developed friction factor correlation from (Ref.~\onlinecite{vangeffelen2021friction}) for micro- and mini-channels with offset strip fins, we have compared it with the actual closure force in the channel, as obtained via DNS. 
The results of this comparison are given in Figure \ref{fig:b_x_t0.02_h0.12_s0.48} for a single channel geometry and four Reynolds numbers.
In this figure, the closure force predicted by the developed friction factor $f_{\text{unit}}$ is given by
\begin{equation}
\label{eq: closure force developed correlations}
\boldsymbol{b} \simeq \boldsymbol{b}_{\text{unit}} \triangleq - \epsilon_{fm} f_{\text{unit}} \frac{2 }{l} \rho_{f}\Vert\langle \boldsymbol{u} \rangle_m \Vert^2  \boldsymbol{e}_s,
\end{equation}
with 
\begin{equation}
f_{\text{unit}} = c_{0} Re_l^{-1} + c_{1}, 
\nonumber
\end{equation}
and
\begin{equation}
\begin{aligned}
  c_{0} &= [23.5 (s/l-t/l)^{-0.83} + 14.9](t/l)^{0.84} (h/l)^{-2} \\
        & \qquad + 13.0 (s/l-t/l)^{-1.69} + 6.0 (h/l)^{-2}, \\
  c_{1} &=  56.5  (s/l-t/l)^{-1.34} (t/l)^{2.94} (h/l)^{-1.08} \\
        & \qquad + 0.0355(s/l-t/l)^{-0.83},\\
\end{aligned}
\label{eq:friction_correlation}
\end{equation}
 whose maximal relative error is below 8\% (Ref.~\onlinecite{vangeffelen2021friction}). Here, $f_{\text{unit}}$ is evaluated based on the local Reynolds number $Re_{l} \left( \boldsymbol{x} \right) \triangleq \rho_{f} \|\langle \boldsymbol{u} \rangle_m \| l / \mu_{f}$, which depends on the local macro-scale velocity $\langle \boldsymbol{u} \rangle_m\left( \boldsymbol{x} \right)$ and its local direction, $\boldsymbol{e}_s \triangleq \langle \boldsymbol{u} \rangle_m /  \Vert \langle \boldsymbol{u} \rangle_m \Vert$.
%%The direction of the closure force  modelled by $f_{\text{unit}}$ is assumed to be anti-parallel to the macro-scale velocity:
%%\begin{equation}
%%\label{eq: direction closure force developed correlations}
%%\frac{\boldsymbol{b} }{\Vert \boldsymbol{b} \Vert}= -\frac{\langle \boldsymbol{u} \rangle_m }{\Vert \langle \boldsymbol{u} \rangle_m \Vert} \,.
%%\end{equation}
The former relationship (\ref{eq: closure force developed correlations}) 
%%and (\ref{eq: direction closure force developed correlations}) 
relies on the assumption that the local closure force is balanced by the macro-scale pressure gradient, i.e. $ \boldsymbol{b} = \nabla{\langle p \rangle_{m}}$, in a similar way as if the flow were periodically developed.
Under that assumption, we can indeed use
the friction factor $f_{\text{unit}}\triangleq \Vert \mathrm{\nabla{P}} \Vert l/(2 \rho_f U^2)$, because
the developed macro-scale closure force effectively agrees with the spatially constant developed pressure gradient $\mathrm{\nabla{P}}$, while the 
the developed macro-scale velocity effectively equals the constant  volume-averaged velocity $\langle \boldsymbol{u} \rangle$, 
(Ref.~\onlinecite{buckinx2015multi}):
\begin{gather}
\langle \boldsymbol{u} \rangle_{m} = \langle \boldsymbol{u} \rangle = U \boldsymbol{e}_{1},  \\
\boldsymbol{f}_{\text{closure}} =  \boldsymbol{b} = \epsilon_{f} \nabla{\langle p \rangle_{m}^{f}} =  \epsilon_{f} \mathrm{\nabla{P}}. 
\label{eq:NavierStokesMS_uniform}
\end{gather}
% Note that the volume-averaged velocity vector $\langle \boldsymbol{u} \rangle$ is defined as 
% \begin{equation}
% \langle \boldsymbol{u} \rangle \triangleq \frac{1}{V_{\text{unit}}} \int_{\boldsymbol{r} \in \Omega_{\text{unit}} \left( \boldsymbol{x} \right)} \boldsymbol{u} \left( \boldsymbol{r} \right) \,d\Omega \left( \boldsymbol{r} \right) \quad \quad \quad \text{with} \quad V_{\text{unit}} = \boldsymbol{l}_{1} \cdot \left( \boldsymbol{l}_{2} \times \boldsymbol{l}_{3} \right). 
% \label{eq:volumeaverage}
% \end{equation}
In principle though, the relationship (\ref{eq: closure force developed correlations}) is only exact over the part of the developed flow region where the flow exhibits both streamwise and transversal periodicity: $\boldsymbol{u} \left( \boldsymbol{x} + \boldsymbol{l}_{j} \right) = \boldsymbol{u} \left( \boldsymbol{x} \right)$ for $j=\{ 1,2 \}$. 
This subregion of $\Omega_{\text{dev}}$, which we denote as $\Omega_{\text{uniform}}$ in this work, is located at a certain distance $l_{\text{sides}} \simeq l_1$ from the channel side walls $\Gamma_{\text{sides}}$
(Ref.~\onlinecite{buckinx2022arxiv}). 
The influence of the side-wall region on the macro-scale flow and the exact extent of $\Omega_{\text{uniform}}$ will be discussed in detail in Section \ref{sec:sidewall}.

According to our DNS results in Figure \ref{fig:b_x_t0.02_h0.12_s0.48}, the developed correlation (\ref{eq:friction_correlation}) is able to capture the actual macro-scale closure force $\boldsymbol{b}$ in $\Omega_{\text{uniform}}$ with a maximum relative error below 5\%, which falls within the correlation accuracy. 
This observation also holds for all the other Reynolds numbers and channel geometries considered in this work.
%%Figure \ref{fig:b_x_t0.02_h0.12_s0.48} illustrates this for one channel geometry and multiple Reynolds numbers at various lateral coordinates $x_{2}$ in the fin array. 

%% entrance region
More interesting is our observation that also in the region of developing flow, $\Omega_{\text{predev}}$, the discrepancies between the actual macro-scale closure force and its prediction based on (\ref{eq: closure force developed correlations}) $\| \boldsymbol{b} - \boldsymbol{b}_{\text{unit}} \| / \| \boldsymbol{b} \|$ are very modest. 
As shown in Figure \ref{fig:b_x_t0.02_h0.12_s0.48}, the discrepancies are virtually negligible, if we look at the component of the closure force along the main flow direction, $b_{1} = \boldsymbol{b} \cdot \boldsymbol{e}_{1}$. 
%%For this, $b_{1}$ is computed by equation (\ref{eq:closureforce}) and compared to the value obtained through the friction factor correlation from (Ref.~\onlinecite{vangeffelen2021friction}) by employing the local Reynolds number $Re_{l} \left( \boldsymbol{x} \right)$. 
More generally, we have found that for all Reynolds numbers and geometrical parameters considered in this work, the friction factor correlation (\ref{eq:friction_correlation}) predicts the actual closure force $b_{1}$ in $\Omega_{\text{predev}}$ with a mean and a maximum error of 2\% and 15\%, respectively. 

%The good correspondence between the developed friction factor and the actual closure force 
The relatively small discrepancies between the actual closure force $\boldsymbol{b}$ and the predicted closure force $\boldsymbol{b}_{\text{unit}}$ in $\Omega_{\text{predev}}$ originate mainly from the assumption of locally periodic flow, but not the presumed balance between the macro-scale closure force and macro-scale pressure gradient itself. 
The reason is that both the macro-scale inertia term $\nabla \boldsymbol{\cdot} (\epsilon_{fm} \langle \boldsymbol{u} \rangle_m \langle \boldsymbol{u} \rangle_m) $ and momentum dispersion source $ \rho_f \nabla \boldsymbol{\cdot} \boldsymbol{M}$ are insignificant in $\Omega_{\text{predev}}$, according to our DNS results, so that the macro-scale momentum equation effectively reduces to $\boldsymbol{b} \simeq \nabla \langle p \rangle_m$ in $\Omega_{\text{predev}}$.
This observation is supported by the length-scale arguments given in the work of Whitaker (Ref.~\onlinecite{whitaker1996forchheimer}).
Besides, our empirical evidence is in line with the results from (Ref.~\onlinecite{buckinx2022arxiv}), which also indicated that for low to moderate Reynolds numbers $(Re_{b} \gtrsim 1)$, the macro-scale momentum dispersion tensor has an insignificant contribution to the total macro-scale closure force ($\| \boldsymbol{b} \| \gg \rho_{f} \| \nabla \cdot \boldsymbol{M} \| $), such that $\boldsymbol{f}_{\text{closure}} \simeq \boldsymbol{b}$. 

Besides these small discrepancies, we remark that the overall spatial variations of $\boldsymbol{b}$ in $\Omega_{\text{predev}}$ remain limited, as they have the same order of magnitude as the variations in the developing macro-scale velocity, which we discussed in Section \ref{sec:onsetdevmacro}. 
%%This is a consequence of the fact that in offset strip fin micro- and mini-channels, Darcy’s law is dominant, such that the pressure gradient scales approximately linear with the macro-scale velocity: $\Vert \nabla \mathrm{P} \Vert \sim \mu_f  \Vert \langle \boldsymbol{u} \rangle \Vert$ (Ref.~\onlinecite{vangeffelen2021friction}). 
Therefore, it can be concluded that the macro-scale closure force, with which the pressure drop over the entire array can be modeled, can be accurately predicted by the developed friction factor correlation in offset strip fin micro- and mini-channels. 

As we will show in the next section, the small discrepancies are completely caused by the onset and thus physics of quasi-periodically developed flow, and hence the eigenvalues and modes that characterize the developing flow in $\Omega_{\text{predev}}$. 
The onset point from which this exponential mode is predominant will be discussed in the next section. 

% We remark that, in this comparison, the local Reynolds number $Re_{l}$ differs from the bulk Reynolds number $Re_{b}$ as a result of the no-slip boundary conditions at the channel side walls, such that $Re_{l} \left( \boldsymbol{x} \right) = Re_{b} \|\langle \boldsymbol{u} \rangle \| / u_{b}$ with $\|\langle \boldsymbol{u} \rangle\| > u_{b}$ in $\Omega_{\text{uniform}}$. 
% conversion used: $b_{1} l_{1} / \left( 2 \rho u_{b}^{2} \right) \simeq 2 f_{\text{unit}} \|\langle \boldsymbol{u} \rangle\|^{2} / u_{b}^{2}$, with $Re_{l} = Re_{b} \|\langle \boldsymbol{u} \rangle\| / u_{b}$. 
% The influence of the side-wall region on the macro-scale flow and the exact extent of $\Omega_{\text{uniform}}$ will be discussed in detail in Section \ref{sec:sidewall}. 

% The small discrepancies are a result of the short development lengths and the small flow angles of attack throughout the offset strip fin array, causing the no-slip force components to show limited variation in the developing macro-scale flow region. 
% % and hence allowing to neglect the transversal macro-scale no-slip force component $b_{2}$. 
% It can be concluded that the macro-scale closure force, with which the pressure drop over the entire array can be modelled, can be accurately predicted by the developed friction factor correlation in offset strip fin micro- and mini-channels. 

% \newpage
% \clearpage

%\newpage
%\clearpage

\section{\label{sec:onsetquasi}Onset of quasi-developed macro-scale flow}

The fact that the flow development length, as well as the deviations from the developed macro-scale velocity and macro-scale pressure gradient, are observed to remain small in offset strip fin micro- and mini-channels, can be explained from the inherent characteristics of the quasi-periodically developed flow regime. 
As we will demonstrate next, this regime is the main mechanism behind the flow development according to our DNS results, as it prevails over almost the entire region of developing macro-scale flow $\Omega_{\text{predev}}$.

\subsection{\label{sec:onsetquasiperiodic}Onset of quasi-periodically developed flow}

In the quasi-periodically developed flow regime, the velocity field $\boldsymbol{u}$ converges asymptotically to the periodically developed velocity field $\boldsymbol{u}^{*}$ along the main flow direction, through a single exponential mode:
\begin{equation}
\boldsymbol{u} \simeq \boldsymbol{u}^{*} + \textbf{U} \exp \left( - \lambda x_{1}  \right) \,.
\label{eq:quasivel}
\end{equation}
The eigenvalue $\lambda$ of this mode is spatially constant, whereas the amplitude $\textbf{U}$ is periodic along the main flow direction: $\textbf{U} \left( \boldsymbol{x} + \boldsymbol{l}_{1} \right) = \textbf{U} \left( \boldsymbol{x} \right)$. 
Both $\lambda$ and $\textbf{U}$ are the solution of the eigenvalue problem presented in (Ref.~\onlinecite{buckinx2023quasiperiodically}). 
% (Ref.~\onlinecite{buckinx2022arxiv})

\begin{figure}[ht!]
\centering
\begin{minipage}{.475\textwidth}
% \centering
\raggedleft
\includegraphics[scale = 1.00]{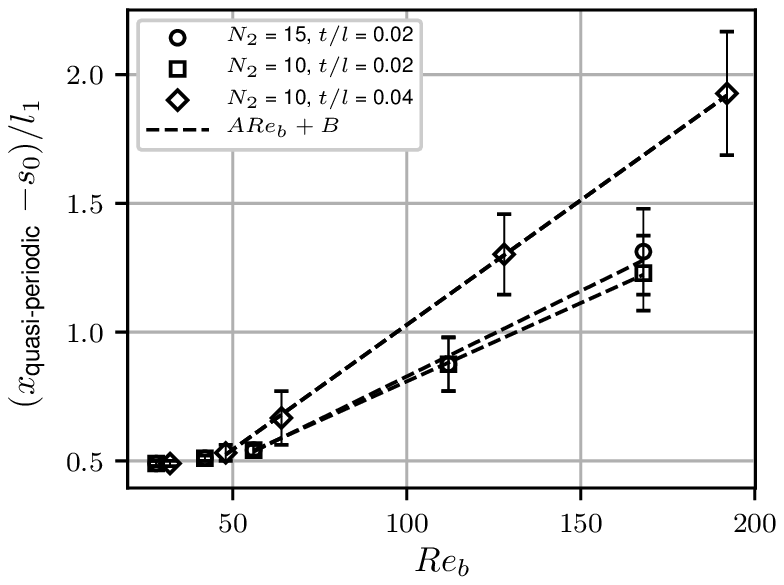}
\caption{\label{fig:onset_RE_t} Influence of the Reynolds number on the onset of the quasi-developed flow, when $N_{1}=20$, $h/l=0.12$, $s/l=0.48$}
\end{minipage}
\hfill
\begin{minipage}{.475\textwidth}
% \centering
\raggedright
% \raggedleft
\includegraphics[scale = 1.00]{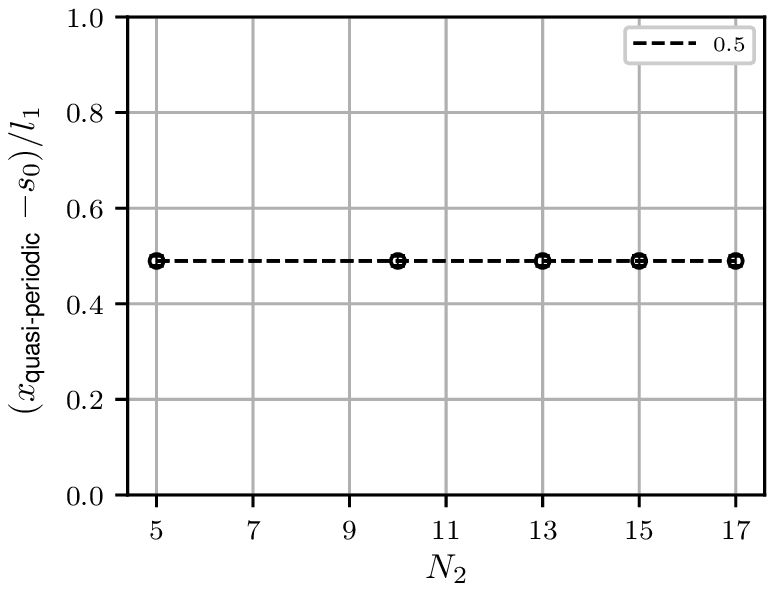}
\caption{\label{fig:onset_N2} Influence of the channel aspect ratio on the onset of the quasi-developed flow, when $Re_{b}=28$, $N_{1}=20$, $h/l=0.12$, $s/l=0.48$, $t/l=0.02$}
\end{minipage}
\end{figure}

%The onset point of the quasi-periodically developed flow regime, $x_{\text{quasi-periodic}}$, which defines the extent of the periodically developed flow region, $x_{1} \in (x_{\text{quasi-periodic}},x_{\text{periodic}})$, is illustrated in Figure xx for a specific geometry and different Reynolds numbers. 
%In this figure, also the mode amplitude $\textbf{U}$ is shown.
The extent of the quasi-periodically developed flow region, $x_{1} \in (x_{\text{quasi-periodic}},x_{\text{periodic}})$, is determined by the onset point of the quasi-periodically developed flow regime, $x_{\text{quasi-periodic}}$. 
Similar to how we defined the onset point of streamwise periodically developed flow, the onset point $x_{\text{quasi-periodic}}$ has been determined as the coordinate $x_{1}$ after which the actual velocity field agrees with expression (\ref{eq:quasivel}) within an accuracy of 1\%. 
%% moved section 
It can be observed in Figure \ref{fig:onset_RE_t} that the onset point $x_{\text{quasi-periodic}}$ scales linearly with the Reynolds number $Re_{b}$.
%%whereas the factor $\epsilon_{0}$ can be considered a constant for each channel geometry. 
In particular, for the data in this figure, we found that $(x_{\text{quasi-periodic}} - s_{0})/l_{1} \simeq 0.00664 Re_{b} + 0.164$ when $\{ N_{2}=15, t/l=0.02 \}$, $(x_{\text{quasi-periodic}} - s_{0})/l_{1} \simeq 0.00611 Re_{b} + 0.199$ when $\{ N_{2}=10, t/l=0.02 \}$, and $(x_{\text{quasi-periodic}} - s_{0})/l_{1} \simeq 0.00969 Re_{b} + 0.0597$ when $\{ N_{2}=10, t/l=0.04 \}$. 
% In particular, for the data in this figure, we found that $(x_{\text{quasi-periodic}} - s_{0})/l_{1} \simeq 0.00186 Re_{b} + 0.164$ when $\{ N_{2}=15, t/l=0.02 \}$, $(x_{\text{quasi-periodic}} - s_{0})/l_{1} \simeq 0.00171 Re_{b} + 0.199$ when $\{ N_{2}=10, t/l=0.02 \}$, and $(x_{\text{quasi-periodic}} - s_{0})/l_{1} \simeq 0.00310 Re_{b} + 0.0597$ when $\{ N_{2}=10, t/l=0.04 \}$. 
The relative uncertainty of these correlations for $x_{\text{quasi-periodic}}$ is below 5\%. 
It should be added that the former correlations for $x_{\text{quasi-periodic}}$ are only valid for $Re_{b} > 48$, since for $Re_{b} \leqslant 48$, the onset point $x_{\text{quasi-periodic}}$ converges towards the first fin row in the array, as illustrated in Figure \ref{fig:onset_RE_t}:  $x_{\text{quasi-periodic}} \simeq s_{0} + 0.5 l_{1}$. 
In that case, the flow can be considered to be quasi-periodically developed from the first fin row of the array. 

Further, the limited data from Figure \ref{fig:onset_RE_t} indicates that the onset point $x_{\text{quasi-periodic}}$ slightly increases when the fin thickness $t/l$ increases. 
Since this trend is opposite to that of the onset point $x_{\text{periodic}}$, this means that the entire extent of the quasi-periodically developed flow region will become smaller in the fin array as $t/l$ increases, although only marginally. 

% \begin{figure}[ht]
% \centering
% \begin{minipage}{.475\textwidth}
% \centering
% \includegraphics[scale = 1.00]{figures/onset_RE_t.eps}
% \caption{\label{fig:onset_RE_t} Influence of the Reynolds number on the onset of quasi-periodically developed flow, when $h/l=0.12$, $s/l=0.48$}
% \end{minipage}
% \end{figure}

In Figure \ref{fig:onset_N2}, the variation of the onset point of quasi-periodically developed flow with the channel aspect ratio is depicted. 
This figure shows that, because of the low Reynolds number selected, the onset of quasi-periodically developed flow coincides virtually with the first fin row in the array for all aspect ratios in the range $N_{2} \in (5,17)$: $x_{\text{quasi-periodic}} \simeq s_{0} + 0.5 l_{1}$. 
% approximately increases linearly with $N_{2}$ , just like the onset point of periodically developed flow, $x_{\text{periodic}}$ (see Figure \ref{fig:xdev_N2}). 
% However, because of the low Reynolds number selected, the linear increase is so small that for all aspect ratios in the range $N_{2} \in (5,17)$, the onset of quasi-periodically developed flow coincides virtually with the first fin row in the array: $x_{\text{quasi-periodic}} \simeq s_{0} + 0.5 l_{1}$. 

\begin{figure}[ht!]
\centering
\begin{minipage}{.475\textwidth}
% \centering
\raggedleft
\includegraphics[scale = 1.00]{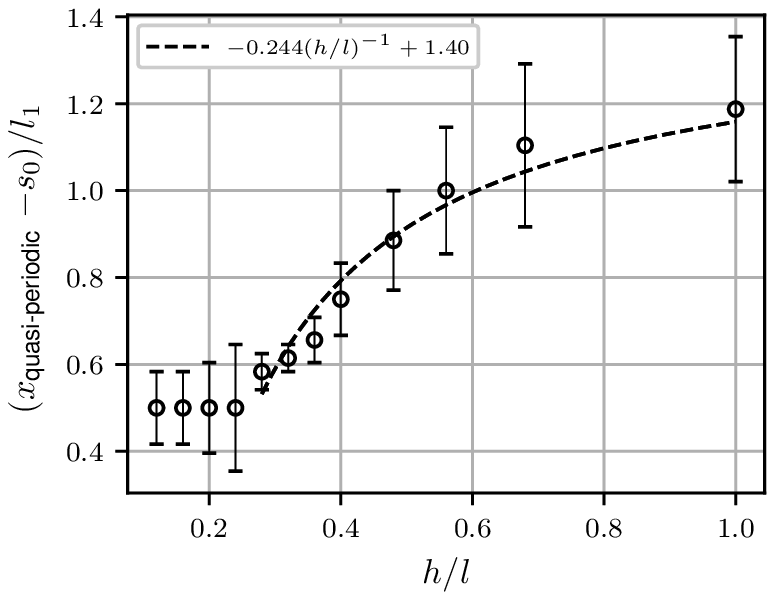}
\caption{\label{fig:onset_h} Influence of the fin height-to-length ratio on the onset of the quasi-developed flow, when $Re_{b} l / (2 L_{3}) = \rho_{f} u_{b} l / \mu_{f} =600$, $N_{1}=20$, $N_{2}=10$, $s/l=0.12$, $t/l=0.02$}
\end{minipage}
\hfill
\begin{minipage}{.475\textwidth}
% \centering
\raggedright
% \raggedleft
\includegraphics[scale = 1.00]{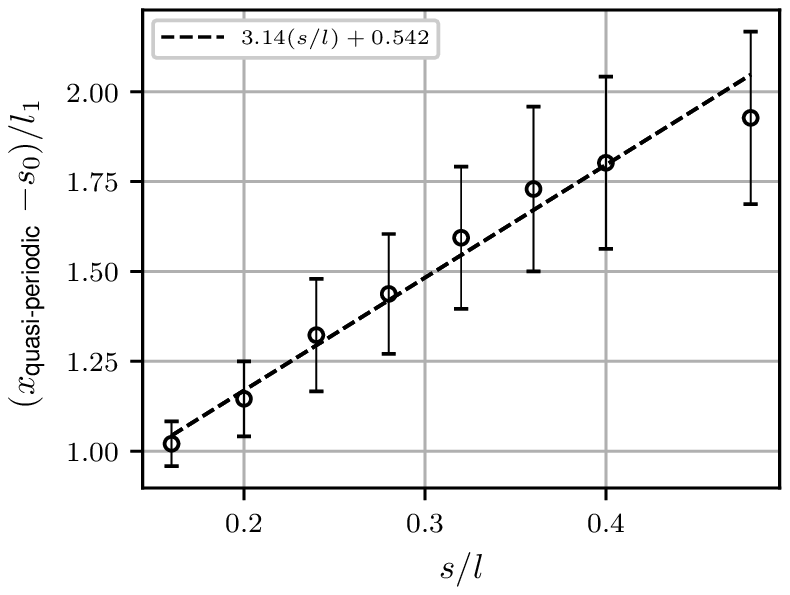}
\caption{\label{fig:onset_s} Influence of the fin pitch-to-length ratio on the onset of the quasi-developed flow, when $Re_{b}=192$, $N_{1}=20$, $N_{2}=10$, $h/l=0.12$, $t/l=0.04$}
\end{minipage}
\end{figure}

In contrast, from the data shown in Figure \ref{fig:onset_h} we found that $x_{\text{quasi-periodic}}$ exhibits a linear relationship with the inverse of the relative fin height $h/l$, just like the onset point of periodically developed flow, $x_{\text{periodic}}$ (see Figure \ref{fig:xdev_h}). 
For the data in Figure \ref{fig:onset_h}, the following correlation was fitted with a maximum error of 8\%: $(x_{\text{quasi-periodic}} - s_{0})/l_{1} \simeq -0.244 (h/l)^{-1} + 1.40$. 
We remark, however, that this correlation for $x_{\text{quasi-periodic}}$ is only valid for $h/l > 0.28$, since for $h/l \leqslant 0.28$, the onset point $x_{\text{quasi-periodic}}$ becomes constant and equal to $ s_{0} + 0.5 l_{1}$. 
It is clear that $x_{\text{quasi-periodic}}$, just like $x_{\text{periodic}}$ becomes independent of the relative height $h/l$ when $h/l$ increases. 
% This can be explained by the fact that when $h/l$ increases, the flow patterns around the fin become more two-dimensional and are influenced less by the top and bottom channel boundaries and their spacing, as was also highlighted for periodically developed flow (Ref.~\onlinecite{vangeffelen2021friction}). 
% This trend with the channel height was also observed for arrays of square cylinders (Ref.~\onlinecite{buckinx2022arxiv}). 

Lastly, the onset point $x_{\text{quasi-periodic}}$ was found to scale linearly with the fin pitch-to-length ratio $s/l$, as depicted in Figure \ref{fig:onset_s}. 
Again, this scaling is similar to the dependence of $x_{\text{periodic}}$ on $s/l$ (see Figure \ref{fig:xdev_s}). 
For instance, within a relative error of 6\%, we obtained the correlation $(x_{\text{quasi-periodic}} - s_{0})/l_{1} \simeq 3.14 (s/l) + 0.542$ for $Re_{b}=192$, $N_{2}=10$, $h/l=0.12$, $t/l=0.04$ and $s/l \in (0.12,0.48)$. 
% through a least-squares fitting approach.

% Lastly, the limited data from Figure \ref{fig:onset_RE_t} indicates that the onset point $x_{\text{quasi-periodic}}$ slightly increases when the fin thickness $t/l$ increases. 
% Since this trend is opposite to that of the onset point $x_{\text{periodic}}$, this means that the entire extent of the quasi-periodically developed flow region will become smaller in the fin array, as $t/l$ increases. 

%%major conclusion missing!!!
According to all our DNS results, the onset point of quasi-periodically developed flow $x_{\text{quasi-periodic}}$ does not exceed the length of two unit cells $2 l_{1}$ from the start of the fin array. 
Therefore, it can be concluded that, in offset strip fin micro- and mini-channels, the flow can be considered quasi-periodically developed almost immediately after the start of the fin array $(x_{\text{quasi-periodic}} \simeq x_{\text{in}})$. 
% for Reynolds number and channel geometries relevant for offset strip fin micro- and mini-channel applications
In comparison, the onset of quasi-periodically developed flow in fin arrays of a higher porosity such as arrays of square cylinders lies further downstream ($x_{\text{quasi-periodic}} \simeq$ 5-15) (Ref.~\onlinecite{buckinx2022arxiv}).
A similar conclusion was already drawn for the onset of periodically developed flow $x_{\text{periodic}}$ in Section \ref{sec:onsetperiodic}. 
% Similar to what was concluded for the onset of periodically developed flow in Section \ref{sec:onsetperiodic}, the onset of quasi-periodically developed flow in offset strip fin micro- and mini-channels lies further upstream in comparison to that in fin arrays of a higher porosity such as arrays of square cylinders (Ref.~\onlinecite{buckinx2022arxiv}). 

\newpage
\clearpage


\newpage
\clearpage

\subsection{\label{sec:eigenvaluesquasiperiodic}Eigenvalues and perturbations for quasi-periodically developed flow}

As the flow can be treated as quasi-periodically developed almost directly after the start of the fin array, the onset of periodically developed flow, as well as the onset of developed macro-scale flow, are in the first place determined by the eigenvalue $\lambda$ of the dominant exponential mode from (\ref{eq:quasivel}). 
In the second place, the onset point $x_{\text{periodic}}$ will also be affected by the magnitude $\Vert \textbf{U} \Vert$ of the dominant mode, and thus the specific inlet conditions of the channel flow.
This follows from the relation between the onset point $x_{\text{periodic}}$ and the eigenvalue $\lambda$,
%between the onset points $x_{\text{periodic}}$ and $x_{\text{quasi-periodic}}$, 
which can be written as
\begin{equation}
%%x_{\text{periodic}} - x_{\text{quasi-periodic}} 
x_{\text{periodic}}
\simeq \frac{1}{\lambda} \ln \left( \frac{\epsilon_{0}}{\epsilon} \right), 
\label{eq:lambda_onset}
\end{equation}
according to equation (\ref{eq:quasivel}) (Ref.~\onlinecite{buckinx2022arxiv,buckinx2023quasiperiodically}).
Here, we have $\epsilon=0.01$, as $\epsilon$ is the criterion used to define the onset of periodically developed flow:  $ \| \boldsymbol{u}_{f} - \boldsymbol{u}^{*}_{f} \| / \| \boldsymbol{u}^{*}_{f} \| \leqslant \epsilon$ for $x_{1} \geqslant x_{\text{periodic}}$.
On the other hand, $\epsilon_{0}$ characterizes the magnitude of the mode amplitude, also called the \textit{perturbation size}, which results from the inlet conditions: 
%%%%%%%%%%%%%%%%%%%%%%%%%%%%%%%%%%%%
%%$\epsilon_{0} = \max \limits_{\boldsymbol{x} \in S} \left( \Vert \textbf{U}  \Vert   /\Vert %%\boldsymbol{u}^{*}_{f} \Vert \right) \exp \left( - \lambda x_{\text{quasi-periodic}} \right) $ 
%%%%%%%%%%%%%%%%%%%%%%%%%%%%%%%%%%%%
$\epsilon_{0} = \max \limits_{\boldsymbol{x} \in S} \left( \Vert \textbf{U}  \Vert   /\Vert \boldsymbol{u}^{*}_{f} \Vert \right) $ with $S \triangleq \{\boldsymbol{x} \vert \, x_1=x_{\text{quasi-periodic}} \}$. 
Hence, $\epsilon_{0}$  determines the peak velocity at the onset point of quasi-periodically developed flow: 
%%%%%%%%%%%%%%%%%%%%%%%%%%%%%%%%%%%%
%%$ \| \boldsymbol{u}_{f} - \boldsymbol{u}^{*}_{f} \| / \| \boldsymbol{u}^{*}_{f} \| \leqslant  %%\epsilon_{0} $ 
%%%%%%%%%%%%%%%%%%%%%%%%%%%%%%%%%%%%
$ \| \boldsymbol{u}_{f} - \boldsymbol{u}^{*}_{f} \| / \| \boldsymbol{u}^{*}_{f} \| \leqslant \epsilon_{0} \exp \left( - \lambda x_{\text{quasi-periodic}} \right)$ 
for $x_{1} \geqslant x_{\text{quasi-periodic}}$. 
Relation (\ref{eq:lambda_onset}) shows that when $x_{\text{quasi-periodic}} \simeq x_{\text{in}} $, the onset point of periodically developed flow will scale in accordance with the eigenvalue $\lambda$, 
%%%%%%%%%%%%%%%%%%%%%%%%%%%%%%%%%%%%
%% i.e. $(x_{\text{periodic}} - x_{\text{in}}) \sim 1/\lambda$, 
%%%%%%%%%%%%%%%%%%%%%%%%%%%%%%%%%%%%
i.e. $x_{\text{periodic}} \sim 1/\lambda$, as long as the perturbation size, and thus the specific inlet conditions, have a rather modest influence on the logarithmic perturbation size $\ln \left( \epsilon_{0}/\epsilon \right)$. 
%%We remark that relation (\ref{eq:lambda_onset}) also shows that 
%%$x_{\text{periodic}} \sim 1/\lambda $ when $x_{\text{quasi-periodic}} > x_{\text{in}} $, if we %%can ignore the influence of the perturbation size on $\ln \left( \epsilon_{0}/\epsilon \right)$.
%%This means that the eigenvalue $\lambda$ can even explain the scaling of the onset point of %%periodically developed flow, if the quasi-periodically regime would start more downstream of the %%channel inlet, provided that the latter regime extends over a large part of the channel.
%%For that reason, the eigenvalue $\lambda$ is more generally of interest than the mode magnitude %%$\Vert \textbf{U} \Vert$ or $\epsilon_{0}$, next to the fact that $\lambda$ is unaffected by the %%inlet conditions.
Furthermore, the eigenvalue $\lambda$ is more generally of interest than the mode magnitude $\Vert \textbf{U} \Vert$ or $\epsilon_{0}$, as it is unaffected by the inlet conditions.

%%\begin{figure}[ht]
%%\centering
%%\begin{minipage}{.475\textwidth}
%%% \centering
%%\raggedleft
%%% \includegraphics[scale = 1.00]{figures/lambda_RE_t.eps}
%%\includegraphics[scale = 1.00]{figures/lambda_inv_RE_t.eps}
%%\caption{\label{fig:lambda_RE_t} Influence of the Reynolds %%number on the eigenvalue of the quasi-developed flow, when %%$h/l=0.12$, $s/l=0.48$}
%%\end{minipage}
%%\hfill
%%\begin{minipage}{.475\textwidth}
%%% \centering
%%\raggedright
%%% \raggedleft
%%\includegraphics[scale = 1.00]{figures/onset_RE_t.eps}
%%\caption{\label{fig:onset_RE_t} Influence of the Reynolds %%number on the onset  of quasi-periodically developed flow, %%when $h/l=0.12$, $s/l=0.48$}
%%\end{minipage}
%%\end{figure}

% \begin{figure}[ht]
% \centering
% \begin{minipage}{.475\textwidth}
% \centering
% \includegraphics[scale = 1.00]{figures/lambda_inv_RE_t.eps}
% \caption{\label{fig:lambda_RE_t} Influence of the Reynolds number on the eigenvalue of the quasi-developed flow, when $h/l=0.12$, $s/l=0.48$}
% \end{minipage}
% \end{figure}

\begin{figure}[ht!]
\centering
\begin{minipage}{.475\textwidth}
% \centering
\raggedleft
\includegraphics[scale = 1.00]{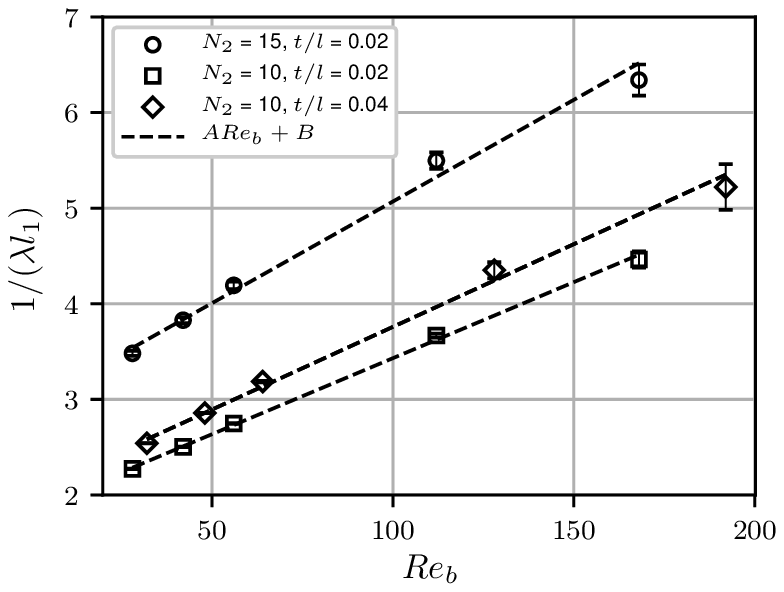}
\caption{\label{fig:lambda_RE_t} Influence of the Reynolds number on the eigenvalue of the quasi-developed flow, when $N_{1}=20$, $h/l=0.12$, $s/l=0.48$ \newline}
\end{minipage}
\hfill
\begin{minipage}{.475\textwidth}
% \centering
\raggedright
% \raggedleft
% \includegraphics[scale = 1.00]{figures/lambda_N2.eps}
\includegraphics[scale = 1.00]{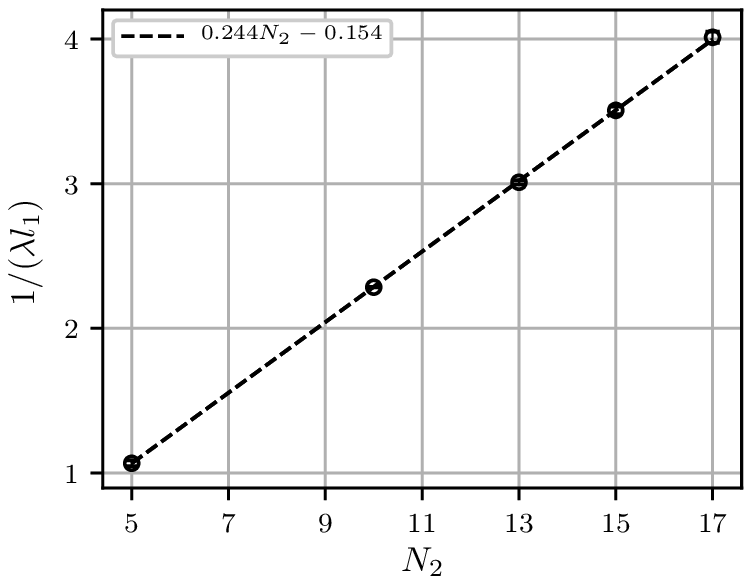}
\caption{\label{fig:lambda_N2} Influence of the channel aspect ratio on the eigenvalue of the quasi-developed flow, when $Re_{b}=28$, $N_{1}=20$, $h/l=0.12$, $s/l=0.48$, $t/l=0.02$}
\end{minipage}
\end{figure}

According to our DNS results, the logarithmic perturbation size $\ln \left( \epsilon_{0}/\epsilon \right)$ is a constant to a first-order approximation when the Reynolds number $Re_{b}$, and thus the mass flow rate, is varied for a fixed channel and array geometry. 
Therefore, our prediction that 
%%%%%%%%%%%%%%%%%%%%%%%%%%%%%%%%%%%%
%% $(x_{\text{periodic}} - x_{\text{in}} ) \sim 1/\lambda \sim 1 + A Re_{b}$
%%%%%%%%%%%%%%%%%%%%%%%%%%%%%%%%%%%%
$x_{\text{periodic}} \sim 1/\lambda \sim A Re_{b} + B$, with $A$ and $B$ some constants, is indeed justified for explaining the influence of the Reynolds number $Re_{b}$ on the onset point $x_{\text{periodic}}$. 
To support this finding, we have illustrated the dependence of the eigenvalue $\lambda$ on $Re_{b}$ in Figure \ref{fig:lambda_RE_t}.
Again, the numerical uncertainties on the eigenvalues obtained via DNS are indicated by uncertainty bars. 
It can be seen that the dimensionless eigenvalue $\lambda l_{1}$ clearly scales inversely linearly with the Reynolds number $Re_{b}$, which explains the linear scaling of the onset point $x_{\text{periodic}}$ with the Reynolds number, as discussed in Section \ref{sec:onsetperiodic}. 
All the calculated eigenvalues in Figure \ref{fig:lambda_RE_t} are captured by an inverse linear relationship with $Re_{b}$ within a maximum relative error of 3\%. 
In particular, for the three channel geometries considered in Figure \ref{fig:lambda_RE_t}, we found for $Re_{b} \in (28,192)$ that $\lambda l_{1} \simeq 1 / \left( 0.0213 Re_{b} + 2.94 \right)$ when $\{ N_{2}=15, t/l=0.02 \}$, $\lambda l_{1} \simeq 1 / \left( 0.0159 Re_{b} + 1.84 \right)$ when $\{ N_{2}=10, t/l=0.02 \}$, and $\lambda l_{1} \simeq 1 / \left( 0.0173 Re_{b} + 2.03 \right)$ when $\{ N_{2}=10, t/l=0.04 \}$. 
% In particular, for the three channel geometries considered in Figure \ref{fig:lambda_RE_t}, we found for $Re_{b} \in (100,600)$ that $\lambda l_{1} \simeq 1 / \left( 0.00596 Re_{b} + 2.94 \right)$ when $\{ N_{2}=15, t/l=0.02 \}$, $\lambda l_{1} \simeq 1 / \left( 0.00445 Re_{b} + 1.84 \right)$ when $\{ N_{2}=10, t/l=0.02 \}$, and $\lambda l_{1} \simeq 1 / \left( 0.00554 Re_{b} + 2.03 \right)$ when $\{ N_{2}=10, t/l=0.04 \}$. 
For these three geometries, we have found respectively that $\epsilon_{0} \simeq 0.17$, $\epsilon_{0} \simeq 0.16$ and $\epsilon_{0} \simeq 0.092$ with a relative error below 8\% over the considered range of $Re_{b}$. 

A similar inversely linear relationship between $\lambda$ and $Re_{b}$ has also been recognized for channels with arrays of square cylinders (Ref.~\onlinecite{buckinx2022arxiv}). 
Nevertheless, for arrays of square cylinders, the coefficient of proportionality $A$ in $1/\lambda \sim A Re_{b} + B$ is up to a factor of three larger such that the eigenvalues are up to two times smaller than for offset strip fin arrays, when $Re_{b} \in (28,192)$. 
This explains further the significantly smaller flow development lengths we observed in Section \ref{sec:onsetperiodic}, in comparison to square cylinder arrays (Ref.~\onlinecite{buckinx2022arxiv}).
For quasi-developed Poisseuille flow in plate channels without solid structures, previous works have found a reciprocal relationship between the first real eigenvalue and the Reynolds number: $1/\lambda \sim A Re_{b}$, at least for Reynolds number above fifty (Refs.~\onlinecite{wilson1969development,sadri2002accurate}). 
The additional constant $B$ in the previous scaling law $1/\lambda \sim A Re_{b} + B$, which determines the asymptotic behavior of the eigenvalue $\lambda$ and onset point $x_{\text{periodic}}$ towards the Stokes flow regime ($Re_{b} \rightarrow 0$), seems therefore only a property of channel flows confined by a solid whose cross-sectional area varies periodically along the main flow direction. 
% $Re_{b} > 50 $
% Nevertheless, in those works, the channel flow was assumed to be two-dimensional such that the no-slip boundary condition at the side walls was not taken into account. 
% Therefore, a direct comparison with the findings in this work is not justified. 

\begin{figure}[ht!]
\includegraphics[scale = 1.00]{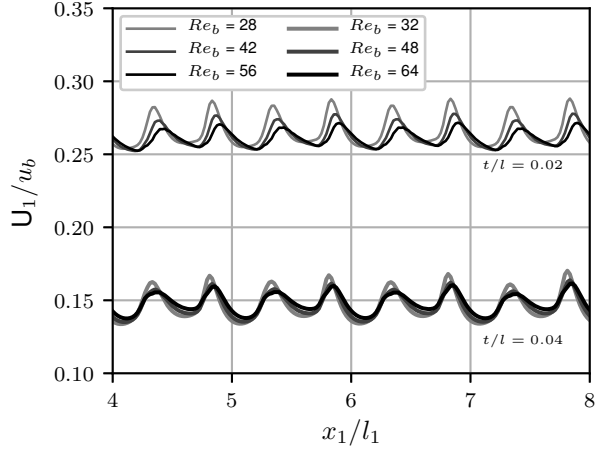}
\caption{\label{fig:Umode_RE_t} Mode amplitude along the channel centerline ($x_{2} = L_{2}/2$, $x_{3} = L_{3}/2$), when $h/l=0.12$, $s/l=0.48$, $s_{0}=l_{1}$, $s_{N}=2.5 l_{1}$, $N_{1}=20$, and $N_{2}=10$}
\end{figure}

Our observation that the logarithmic perturbation size $\ln \left( \epsilon_{0}/\epsilon \right)$ in (\ref{eq:lambda_onset}) can be considered constant over a wide range of Reynolds numbers for each channel geometry, is a consequence of our assumption that the shape of the velocity profile $\boldsymbol{u}_{\text{in}}/u_b$ at the channel inlet (see Section \ref{sec:eqs}) is not affected by the Reynolds number, or the mass flow rate. 
% Although this is evidently an idealization of the inlet conditions, in reality, it is motivated by the empirical evidence of ...
The implication of this assumption is that also the envelope of the dimensionless velocity field $\boldsymbol{u}/u_b$ and the dimensionless peak velocity amplitude $ \max \limits_{\boldsymbol{x} \in S} (\Vert \textbf{U}  \Vert / \Vert \boldsymbol{u}^{\star}  \Vert)$ in the entrance region are barely affected by the Reynolds number. 
This can be understood from Figure \ref{fig:Umode_RE_t}, which shows that the dimensionless mode amplitude (or perturbation size) U$_{1}/u_b $ decreases just very slightly when the Reynolds number $Re_{b}$ is increased, while also $\boldsymbol{u}^{\star} \sim u_b$. 
Given that $\ln \left( \epsilon_{0}/\epsilon \right)$ is a logarithmic function of the dimensionless mode amplitude  $\Vert \textbf{U}  \Vert/u_b $, its dependence on the Reynolds number is thus very weak. 

Further, the limited data considered in Figure \ref{fig:lambda_RE_t} indicates that the scaling of the onset point $x_{\text{periodic}}$ with the fin thickness $t/l$ is even predominantly determined by the magnitude of the mode amplitude, and only secondarily affected by the mode eigenvalue. 
After all, one would expect from the eigenvalues in Figure \ref{fig:lambda_RE_t}, which decrease with increasing $t/l$, that also the onset point $x_{\text{periodic}}$ would move further downstream in the fin array for larger $t/l$ values. 
Nevertheless, as Figure \ref{fig:Umode_RE_t} shows, the mode magnitude decreases more strongly with increasing $t/l$, so that it overcomes the trend imposed by the eigenvalue, and causes the onset point $x_{\text{periodic}}$ to move upstream in the fin array, as shown earlier in Figure \ref{fig:xdev_RE_t}. 

The logarithmic perturbation size is only marginally affected by the channel aspect ratio according to our DNS results. 
Therefore, the eigenvalue $\lambda$ is again responsible for the observed linear scaling of the onset point $x_{\text{periodic}}$ with $N_2$, which was shown in Figure \ref{fig:xdev_N2}. 
Figure \ref{fig:lambda_N2} shows that the reciprocal eigenvalue $1/\lambda$ indeed increases linearly with the transversal number of unit cells $N_{2}$ of the channel, in such a manner that the product $\lambda x_{\text{periodic}}$ is approximately constant.
For example, the correlation $\lambda l_{1} \simeq 1 / ( 0.244 N_{2} - 0.154  )$ matches the data from this work for $Re_{b}=28$, $h/l=0.12$, $s/l=0.48$, $t/l=0.02$ and $N_{2} \in (5,17)$ with an accuracy of 3\%. 
Additionally, for this data, we found that $\epsilon_{0} \simeq 0.00213 N_{2} + 0.158$ within an error of 1\%, such that the factor $\ln \left( \epsilon_{0}/\epsilon \right)$ only varies up to 2\% with $N_{2}$. 
%%% EXPLANATION MISSING

\begin{figure}[ht!]
\centering
\begin{minipage}{.475\textwidth}
% \centering
\raggedleft
\includegraphics[scale = 1.00]{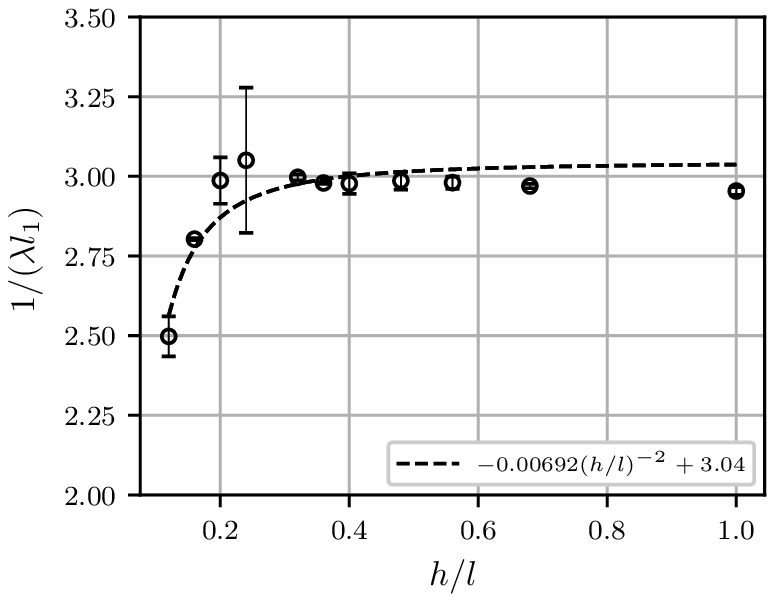}
\caption{\label{fig:lambda_h} Influence of the fin height-to-length ratio on the eigenvalue of the quasi-developed flow, when $Re_{b} l / (2 L_{3}) = \rho_{f} u_{b} l / \mu_{f} =600$, $N_{1}=20$, $N_{2}=10$, $s/l=0.12$, $t/l=0.02$}
\end{minipage}
\hfill
\begin{minipage}{.475\textwidth}
% \centering
\raggedright
% \raggedleft
% \includegraphics[scale = 1.00]{figures/lambda_s.eps}
\includegraphics[scale = 1.00]{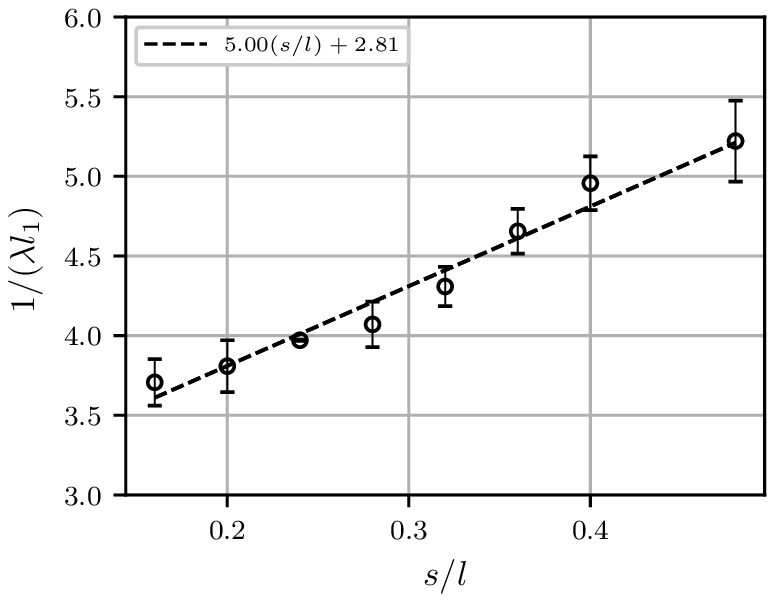}
\caption{\label{fig:lambda_s} Influence of the fin pitch-to-length ratio on the eigenvalue of the quasi-developed flow, when $Re_{b}=192$, $N_{1}=20$, $N_{2}=10$, $h/l=0.12$, $t/l=0.04$}
\end{minipage}
\end{figure}

In contrast, the scaling of the onset point $x_{\text{periodic}}$ with the relative fin height $h/l$ cannot be explained solely on the grounds of the eigenvalue $\lambda$. 
Our DNS results from Figure \ref{fig:lambda_h} indicate that $1/\lambda$ is only weakly influenced by $h/l$ through a linear relation with $(h/l)^{-2}$, whereas $x_{\text{periodic}}$ is strongly dependent on $h/l$ through a linear relation with $(h/l)^{-1}$. 
The reason is that the magnitude of the mode amplitude greatly increases for increasing values of the fin height-to-length ratio $h/l$. 
% when the Reynolds number is kept constant. 
% The reason is that the magnitude of the mode amplitude is strongly affected by any change in the fin height-to-length ratio $h/l$ when the Reynolds number is kept constant. 
% the Reynolds number based on the fin length $Re_l$ is kept constant. 
% Physically, this means that/can be understood from ...
For the conditions illustrated in Figure \ref{fig:lambda_h}, that is for $h/l \in (0.12,1)$, the following correlations were fitted with a maximum error of 6\% and 10\% respectively: $\lambda l_{1} \simeq 1 / (0.00692 (h/l)^{-2} + 3.04 )$ and $\epsilon_{0} \simeq 0.626 (h/l)/ \left( (h/l) + 2.56 \right) + 0.00144$. 
% 0.00549 (h/l) + 0.203
So, the logarithmic perturbation size $\ln \left( \epsilon_{0}/\epsilon \right)$ exhibits a variation of more than 40\% for $h/l \in (0.12,1)$. 
These correlations further indicate that $\lambda$ and $\epsilon_{0}$, and hence $x_{\text{periodic}}$ become independent of the height $h/l$ when $h/l$ increases.

Finally, also the fin pitch-to-length ratio $s/l$ has a significant influence on the magnitude of the mode amplitude in the quasi-periodically developed flow regime, such that the scaling of the onset point $x_{\text{periodic}}$ is not solely determined by the relation between $s/l$ and the eigenvalue $\lambda$. 
% although the scaling of the onset point $x_{\text{periodic}}$ is still mostly determined by the relation between $s/l$ and the eigenvalue $\lambda$. 
More specifically, the inverse of the eigenvalue $1/ \lambda$ scales linearly with the fin pitch-to-length ratio $s/l$, just as the onset point $x_{\text{periodic}}$. 
In particular, within a relative error of 6\%, we have $\lambda l_{1} \simeq 1/( 5.00 (s/l) + 2.81 ) $ for $Re_{b}=192$, $N_{2}=10$, $h/l=0.12$, $t/l=0.04$ and $s/l \in (0.12,0.48)$, as shown in Figure \ref{fig:lambda_s}. 
Still, the perturbation size $\epsilon_{0}$ varies nearly quadratically with fin pitch-to-length ratio $s/l$: $\epsilon_{0} \simeq 0.427 (s/l)^{2}$, within an error of 10\%. 
Therefore, also $\ln \left( \epsilon_{0}/\epsilon \right)$ increases as much as with a factor of ten over the range $s/l \in (0.12,0.48)$. 
Hence, both the behavior of the eigenvalue and perturbation size contribute to the fact that the onset point of periodically developed flow moves significantly further downstream when $s/l$ increases, as we showed in Section \ref{sec:onsetperiodic}.

To summarize, the previous observations show that, in essence, the relatively small flow development lengths in micro- and mini-channels with arrays of offset strip fins can be attributed to the large eigenvalues and small perturbation sizes characterizing the quasi-periodically developed flow, in the region of developing flow. 

\newpage
\clearpage


%%% Original text with epsilon_0 values as a function of Re
% Additionally, in this work, it has been observed that the onset $x_{\text{quasi-periodic}}$ scales linearly with $Re_{b}$, whereas the factor $\epsilon_{0}$ can be considered a constant for each channel geometry. 
% Particularly, we found that for $h/l=0.12$, $s/l=0.48$ $x_{\text{quasi-periodic}} \simeq 0.00186 Re_{b} + 0.164$ and $\epsilon_{0} \simeq 0.08$ when $\{ N_{2}=15, t/l=0.02 \}$, $x_{\text{quasi-periodic}} \simeq 0.00171 Re_{b} + 0.199$ and $\epsilon_{0} \simeq 0.05$ when $\{ N_{2}=10, t/l=0.02 \}$, and $x_{\text{quasi-periodic}} \simeq 0.00310 Re_{b} + 0.0597$ and $\epsilon_{0} \simeq 0.03$ when $\{ N_{2}=10, t/l=0.04 \}$. 

% \newpage
% \clearpage

\subsection{\label{sec:onsetquasidevmacro} Region of quasi-developed macro-scale flow}

Due to the fact that the flow can be treated as quasi-periodically developed very shortly after the start of the fin array, the macro-scale velocity field in the channel is nearly completely described by two terms:
\begin{equation}
\langle \boldsymbol{u} \rangle_{m} \simeq \boldsymbol{U}_{\text{dev}} + \langle \textbf{U} \rangle_{m} \exp \left( - \lambda x_{1}  \right).  
\label{eq:quasivelMS}
\end{equation}
The first term is the developed velocity profile $\boldsymbol{U}_{\text{dev}}$ in $\Omega_{\text{dev}}$, while the second term contains the macro-scale velocity mode $\langle \textbf{U} \rangle_{m} $.
Both depend only on the transversal coordinate $x_2$, but not the streamwise coordinate $x_{1}$. 
Strictly speaking, the decomposition (\ref{eq:quasivelMS}) is only valid one unit cell after the onset of quasi-periodically developed flow, so that the onset point of quasi-developed macro-scale flow is given by $x_{\text{quasi-dev}} = x_{\text{quasi-periodic}} + l_{1}$. 

\begin{figure}[ht]
\centering
\begin{minipage}{.475\textwidth}
% \centering
\raggedleft
\includegraphics[scale = 1.00]{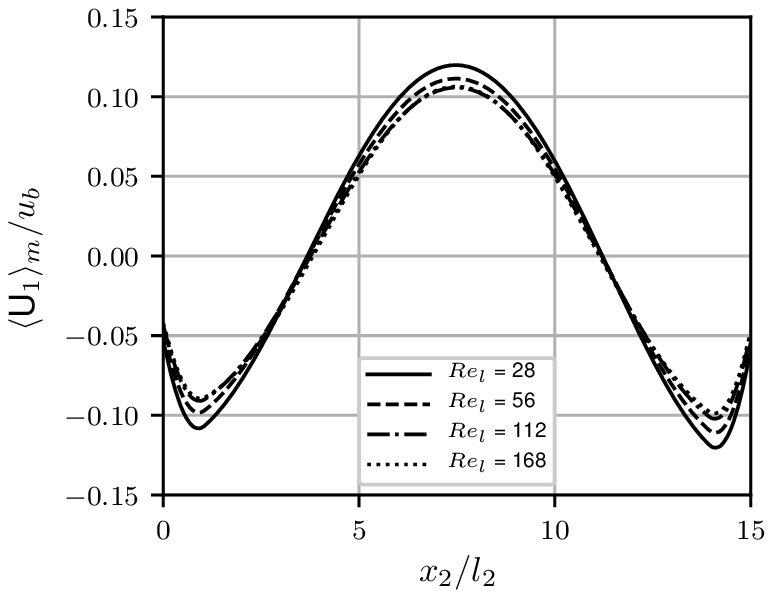}
\caption{\label{fig:U0_profile} Influence of the Reynolds number on the streamwise macro-scale velocity mode for quasi-developed flow, when $N_{2}=15$, $h/l=0.12$, $s/l=0.48$, $t/l=0.02$}
\end{minipage}
\hfill
\begin{minipage}{.475\textwidth}
% \centering
\raggedright
% \raggedleft
\includegraphics[scale = 1.00]{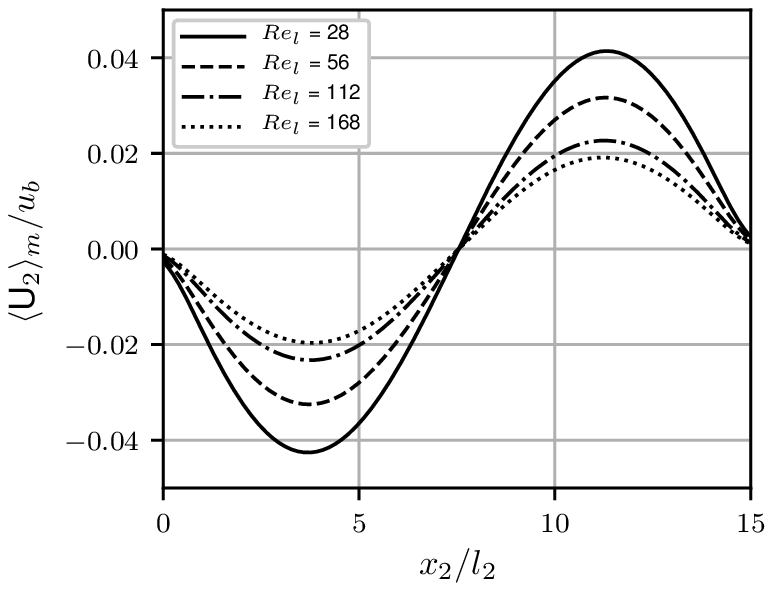}
\caption{\label{fig:U1_profile} Influence of the Reynolds number on the transversal macro-scale velocity mode for quasi-developed flow, when $N_{2}=15$, $h/l=0.12$, $s/l=0.48$, $t/l=0.02$}
\end{minipage}
\end{figure}

The macro-scale velocity modes which occur in micro- and mini-channels with offset strip fin arrays are illustrated for one single channel geometry in Figures \ref{fig:U0_profile} and \ref{fig:U1_profile}. 
Although the shape of the macro-scale mode components $\langle \text{U}_{1} \rangle_{m}$ and $\langle \text{U}_{2} \rangle_{m}$ is only universal for a given Reynolds number and array geometry, it can be seen that the influence of the Reynolds number on their shape is very small. 
In addition, also their absolute magnitude changes insignificantly with the Reynolds number. 
The limited dependence of the preceding modes on the Reynolds number is of course inherited from the original modes $\text{U}_{1} $ and $\text{U}_{2}$, and hence the perturbation size $\epsilon_{0}$, before spatial averaging. 
Therefore, we can represent all the macro-scale velocity modes for a single geometry by a single reference mode $\text{U}_{\text{ref}}$. 
For the geometry selected in Figures \ref{fig:U0_profile} and \ref{fig:U1_profile}, we have for instance
\begin{equation}
\text{U}_{\text{ref}} \triangleq u_b \sin \left( \frac{2 \pi}{L_{2}} (x_2 - L_{2}/4) \right), 
\end{equation}
in $\Omega_{\text{predev}} \setminus \Omega_{\text{sides}}$. 
Based on this reference mode, the modes for all Reynolds numbers in the same figures can be correlated with a maximum error of 3\% as follows:
$\langle \text{U}_{1} \rangle_{m} / \text{U}_{\text{ref}} \simeq 0.49 Re_{b}^{-1} + 0.103$ when $\{ N_{2}=15, t/l=0.02 \}$, $\langle \text{U}_{1} \rangle_{m} / \text{U}_{\text{ref}} \simeq 0.0254 Re_{b}^{-1} + 0.0855$ when $\{ N_{2}=10, t/l=0.02 \}$, and $\langle \text{U}_{1} \rangle_{m} / \text{U}_{\text{ref}} \simeq 0.0562 Re_{b}^{-1} + 0.0531$ when $\{ N_{2}=10, t/l=0.04 \}$. 
% Reason of large difference in proportionality coefficient is the limited dependence on Re over the considered range
% $\langle \text{U}_{1} \rangle_{m} / \text{U}_{\text{ref}} \simeq 1.75 Re_{b}^{-1} + 0.103$ when $\{ N_{2}=15, t/l=0.02 \}$, $\langle \text{U}_{1} \rangle_{m} / \text{U}_{\text{ref}} \simeq 0.0907 Re_{b}^{-1} + 0.0855$ when $\{ N_{2}=10, t/l=0.02 \}$, and $\langle \text{U}_{1} \rangle_{m} / \text{U}_{\text{ref}} \simeq 0.176 Re_{b}^{-1} + 0.0531$ when $\{ N_{2}=10, t/l=0.04 \}$. 
The transversal macro-scale velocity mode $\langle \text{U}_{2} \rangle_{m}$ can be computed through integration of $\langle \text{U}_{1} \rangle_{m}$ over $x_{2}$, given that $\nabla \cdot \textbf{U} = \lambda \text{U}_{1}$, as derived in (Ref.~\onlinecite{buckinx2022arxiv}). 
% The linear dependence of the preceding modes on the inverse Reynolds number is of course inherited from the original modes $\text{U}_{1} $ and $\text{U}_{2}$ before spatial averaging.

The former mode correlations as a function of $Re_{b}$ are in agreement with those found for quasi-developed Poisseuille flow in channels without fins at Reynolds numbers below 500 (Ref.~\onlinecite{sadri1997channel}), as well as those found for quasi-developed flow in channels with arrays of square cylinders at Reynolds numbers between 25 and 200 (Ref.~\onlinecite{buckinx2022arxiv}). 
Nevertheless, the magnitude of the macro-scale velocity modes observed here for offset strip fin arrays is approximately one order of magnitude smaller than for square cylinder arrays. 
In addition, our DNS data reveals that the previous macro-scale modes $\langle \text{U}_{1} \rangle_{m} / \text{U}_{\text{ref}}(x_2)$ do not vary more than 5\% with the Reynolds number for $Re_{b} \in (28,192)$, whereas for arrays of square cylinders, these modes vary around 20\% with $Re_{b}$ for the same boundary conditions. 
%Why is this the case??

%%% Paragraphs/information missing:

% For all Reynolds numbers and geometrical parameters considered in this work, the perturbation sizes in the entrance region of the channel remain limited with respect to the developed macro-scale flow for offset strip fin micro- and mini-channels. 
% This is a direct result of both the large eigenvalues and small amplitudes of the dominant exponential mode in offset strip fin arrays. 

The small magnitudes of the macro-scale modes could have already been observed from the developing macro-scale velocity profiles illustrated earlier in Figure \ref{fig:msvel}. 
% The same conclusion could have already been drawn from the developing macro-scale velocity profiles illustrated earlier in Figure \ref{fig:msvel}. 
Yet, the macro-scale velocity modes give a more general and concise picture of the developing flow. 
Moreover, their small magnitude explains directly why the deviations from the developed macro-scale closure force remain limited even when the flow is still developing so that we can rely on the developed friction factor $f_{\text{unit}}$ to model the closure force over nearly the entire channel, as shown in Figure \ref{fig:b_x_t0.02_h0.12_s0.48}. 
The reason is that the true friction factor $f$ which governs the closure force in the quasi-developed flow region is given by
\begin{equation}
f \simeq f_{\text{unit}} + 
\frac{l}{2\rho_f  \epsilon_{fm} \Vert \langle \boldsymbol{u} \rangle_m \Vert  } \mu_{f} \boldsymbol{\mathrm{K}}^{-1} \boldsymbol{\cdot} \boldsymbol{\zeta} 
% \boldsymbol{\cdot} \boldsymbol{\xi}^{-1} 
\boldsymbol{\cdot} \boldsymbol{e}_s  \exp(-\lambda x_1)\,,
\end{equation}
in $\Omega_{\text{predev}} \setminus \Omega_{\text{sides}}$, 
as deduced by (Ref.~\onlinecite{buckinx2022arxiv}). 
Here, $\boldsymbol{\zeta}$ denotes the transformation tensor mapping $U e_{1}$ onto $\langle \textbf{U} \rangle_{m}$ and $\boldsymbol{\mathrm{K}}$ the permeability tensor representing the additional resistance resulting from the macro-scale velocity mode $\langle \textbf{U} \rangle_{m}$ (Ref.~\onlinecite{buckinx2022arxiv}). 
As such, the true friction factor $f$ virtually equals $f_{\text{unit}}$ for small (tensorial) velocity perturbations $\boldsymbol{\zeta}$
% , since $\boldsymbol{\xi}$ corresponds to the identity tensor (at least outside the side-wall region)
, while the additional permeability tensor $\boldsymbol{\mathrm{K}}$ only affects the shape of the resulting closure force modes.

\newpage
\clearpage

\section{\label{sec:sidewall}Influence of the side-wall region on the macro-scale flow}

In accordance with (\ref{eq:quasivelMS}), a complete specification of the macro-scale velocity field in the developed and quasi-developed regions, requires full knowledge of the developed macro-scale velocity profile $\boldsymbol{U}_{\text{dev}}$, next to the previously discussed modes. 
Although it has been common in the literature to assume that this profile is uniform and equal to the bulk velocity, $\boldsymbol{U}_{\text{dev}} \simeq u_b \boldsymbol{e}_1$ (Refs.~\onlinecite{zhou2011numerical,kim2011correlations}), this will not be exactly the case, due to the presence of a side-wall region, as we illustrated already in Figure \ref{fig:msvel}(a). 
In the side-wall region, the flow is no longer periodic along the lateral direction $\boldsymbol{e}_{2}$, as it is in the core of the developed flow region $\Omega_{\text{dev}}$, due to the no-slip boundary condition at the side walls of the channel. 
On a macro-scale level, the viscous stresses near the side walls $\Gamma_{\text{sides}}$ will cause the macro-scale velocity to decrease towards the side walls.
In addition, there occurs a porosity gradient in the side-wall region, along the lateral direction $\boldsymbol{e}_{2}$. 
Therefore, we will now characterize the influence of this side-wall region on the macro-scale flow. 

% and an increase of the local macro-scale closure force. 

\subsection{\label{sec:sidewallmacrovel} Macro-scale velocity profile in the side-wall region}

Our first observation is that the side-wall region practically extends over the width of a single unit cell in the lateral direction $l_{2}$, i.e. the distance over which the lateral porosity gradient occurs. 
The side-wall region $\Omega_{\text{sides}}$ therefore practically corresponds to $x_{2} \in (0,l_{2}) \cup (L_{2}-l_{2},L_{2}) $. 
The reason is that the distance from the side walls $l_{\text{sides}}$ over which the flow in $\Omega_{\text{dev}}$ loses its transversal periodicity, is actually even smaller than the unit-cell width $l_{2}$. 
At least, we found that $l_{\text{sides}} < l_{2}$ for all the micro- and mini-channels with offset strip fins investigated in this work. 
This is supported by the evidence in the study of channels with arrays of in-line square cylinders (Ref.~\onlinecite{buckinx2022arxiv}). 
As a result, the region where the macro-scale velocity $\boldsymbol{U}_{\text{dev}}$ is uniform, coincides with the region of constant porosity and is given by $\Omega_{\text{uniform}} =\{ \boldsymbol{x} \in \Omega | x_{1} \in (x_{\text{dev}},x_{\text{out}}), x_{2} \in (l_{2}, L_{2} - l_{2} ) \}$. 

\begin{figure}[ht]
\includegraphics[scale=1.0]{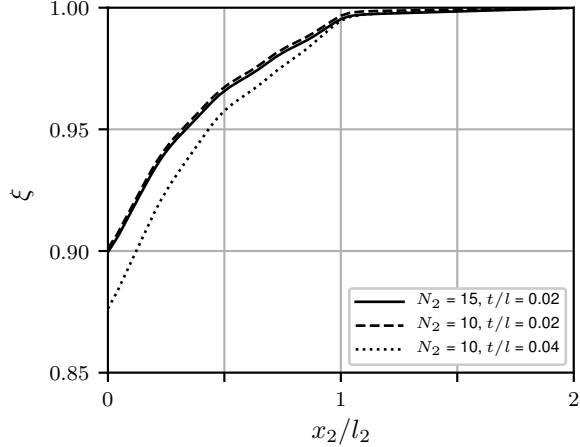}
\caption{\label{fig:xi_y} Macro-scale velocity profile in the side-wall region, when $Re_{b} l / (2 L_{3}) = \rho_{f} u_{b} l / \mu_{f} =600$, $h/l=0.12$, $s/l=0.48$}
\end{figure}

Our second observation is that the macro-scale velocity profile in the side-wall region is quadratic in good approximation. 
This can be seen from Figure \ref{fig:xi_y}, which illustrates the shape of the macro-scale velocity profile $\xi$ for three offset strip fin channel geometries. 
This profile is defined as $\xi \left( x_{2} \right) = U'_{dev} \left( x_{2} \right) /U'$, such that it maps the local macro-scale velocity in the side-wall region $U'_{dev} \left( x_{2} \right) \triangleq \epsilon_{fm}^{-1} U_{dev} \left( x_{2} \right)$ to the uniform macro-scale velocity $U' \triangleq \epsilon_{f}^{-1} U$. 
Note that $\epsilon_{fm}$ is a function of the coordinate $x_{2}$, whereas $\epsilon_{f}$ is the spatially constant porosity in $\Omega_{\text{uniform}}$. 
Based on all our DNS results, the following quadratic approximation of the macro-scale velocity profile holds within a relative error of 4\%:
% \begin{equation}
% \xi \left( x_{2} \right) \simeq
% \begin{cases}
% \frac{x_{2} + l_{\text{slip}}}{l_{2} + l_{\text{slip}}}           & \text{for } x_{2} \in (0, l_{2}), \\
% 1                                                   & \text{for } x_{2} \in (l_{2}, L_{2}-l_{2}), \\
% \frac{L_{2} - x_{2} + l'_{\text{slip}}}{l_{2} + l'_{\text{slip}}} & \text{for } x_{2} \in (L_{2}-l_{2}, L_{2}). 
% \end{cases}
% \end{equation}
\begin{equation}
\xi \left( x_{2} \right) \simeq
\begin{cases}
\frac{1}{1 + 2 (l_{\text{slip}} / l_{2})} \left[ 2 \left( \frac{l_{\text{slip}}}{l_{2}} \right) + 2 \left( \frac{x_{2}}{l_{2}} \right) -\left( \frac{x_{2}}{l_{2}} \right)^{2} \right]                 & \text{for } x_{2} \in (0, l_{2}), \\
1 & \text{for } x_{2} \in (l_{2}, L_{2}-l_{2}), \\
\frac{1}{1 + 2 (l'_{\text{slip}} / l_{2})} \left[ 2 \left( \frac{l'_{\text{slip}}}{l_{2}} \right) + 2 \left( \frac{L_{2} - x_{2}}{l_{2}} \right) -\left( \frac{L_{2} - x_{2}}{l_{2}} \right)^{2} \right] & \text{for } x_{2} \in (L_{2}-l_{2}, L_{2}). 
\end{cases}
\label{eq:xi}
\end{equation}
% We remark that $\xi$ can also be recognized in the spacing of the grey lines in Figure \ref{fig:msvel}
This approximation for the velocity profile $\xi$ is fully determined by the so-called slip lengths $l_{\text{slip}}$ and $l'_{\text{slip}}$ on each side of the channel (Ref.~\onlinecite{buckinx2022arxiv}). 
The latter can be taken equal, i.e. $l_{\text{slip}} \simeq l'_{\text{slip}}$, as their relative difference is smaller than their numerical uncertainty of 6\%, even though the flow, as well as the geometry of the offset strip array, are not symmetric with respect to the plane $x_2=L_2/2$ of the channel. 

\begin{figure}[ht!]
\centering
\begin{minipage}{.475\textwidth}
% \centering
\raggedleft
\includegraphics[scale = 1.0]{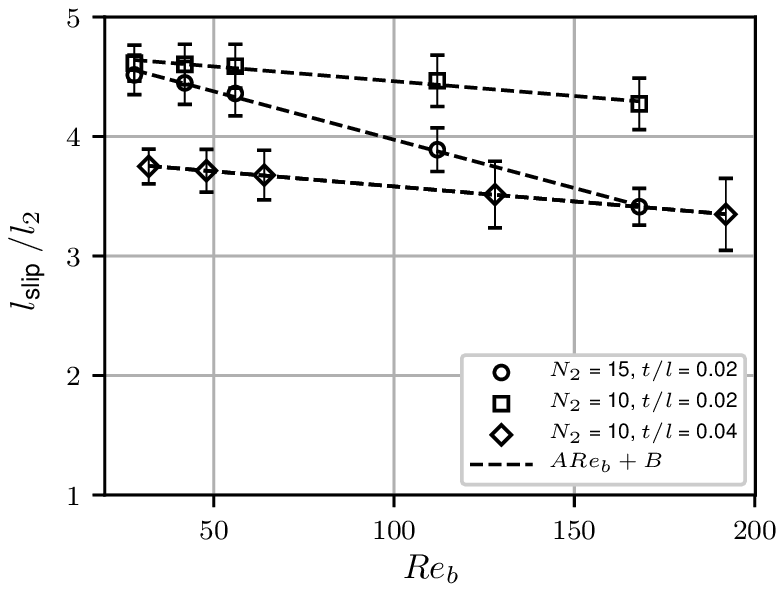}
\caption{\label{fig:lslip_RE_t} Influence of the Reynolds number on the macro-scale slip length, when $N_{1}=20$, $h/l=0.12$, $s/l=0.48$}
\end{minipage}
\hfill
\begin{minipage}{.475\textwidth}
% \centering
\raggedright
% \raggedleft
\includegraphics[scale = 1.0]{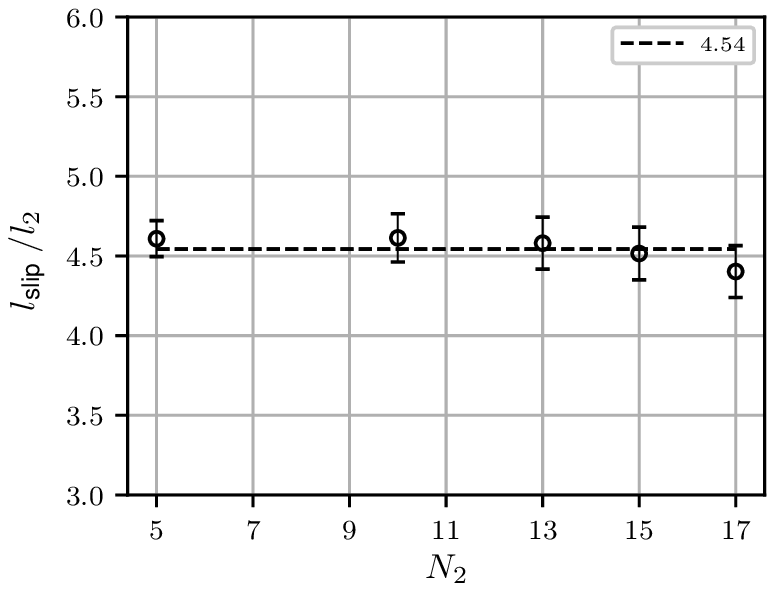}
\caption{\label{fig:lslip_N2} Influence of the channel aspect ratio on the macro-scale slip length, when $Re_{b}=28$, $N_{1}=20$, $h/l=0.12$, $s/l=0.48$, $t/l=0.02$}
\end{minipage}
\end{figure}

From our DNS results in Figure \ref{fig:lslip_RE_t}, we learn that the macro-scale velocity profile, and thus both the slip lengths are just slightly affected by the Reynolds number. 
This observation points towards a similarity with developed (Poiseuille) flow in channels without fins, which is characterized by a Reynolds-number independent velocity profile. 
The limited influence of the Reynolds number on the slip length in the side-wall region can be accurately predicted by a linear relationship based on our DNS data.
Specifically, for an offset strip fin channel with $h/l=0.12$, $s/l=0.48$ and $Re_{b} \in (28,192)$, the following correlations can be fitted through the slip length data, with discrepancies less than 1\%: $l_{\text{slip}} / l_{2} \simeq -0.000811 Re_{b} + 4.78$ when $\{ N_{2}=15, t/l=0.02 \}$, $l_{\text{slip}} / l_{2} \simeq -0.00247 Re_{b} + 4.71$ when $\{ N_{2}=10, t/l=0.02 \}$, and $l_{\text{slip}} / l_{2} \simeq -0.00252 Re_{b} + 3.83$ when $\{ N_{2}=10, t/l=0.04 \}$. 
% Specifically, for an offset strip fin channel with $h/l=0.12$, $s/l=0.48$ and $Re_{b} \in (100,600)$, the following correlations can be fitted through the slip length data, with discrepancies less than 1\%: $l_{\text{slip}} / l_{2} \simeq -0.000227 Re_{b} + 4.78$ when $\{ N_{2}=15, t/l=0.02 \}$, $l_{\text{slip}} / l_{2} \simeq -0.000692 Re_{b} + 4.71$ when $\{ N_{2}=10, t/l=0.02 \}$, and $l_{\text{slip}} / l_{2} \simeq -0.000807 Re_{b} + 3.83$ when $\{ N_{2}=10, t/l=0.04 \}$. 
These correlations show that the relative variations of $l_{\text{slip}}$ with $Re_{b}$ remain below 20\% for each geometry considered in this work. 
As such, the dependence of $l_{\text{slip}}$ on $Re_{b}$ length is still more pronounced than it is for channels with arrays of square cylinders.
For the latter fin geometry, the slip length $l_{\text{slip}}$ is virtually constant for $Re_{b} \in (25,300)$ (Ref.~\onlinecite{buckinx2022arxiv}). 

In contrast, these correlations show that the slip length strongly decreases when the fin thickness-to-length ratio $t/l$ increases.
%as was observed in (Ref.~\onlinecite{buckinx2022arxiv}). 
The explanation is that a larger fin thickness induces a larger lateral velocity $u_{2}$ in the fin array. 
In order to satisfy the no-slip condition at the channel side walls, the lateral periodicity of the flow must thus be interrupted over a larger distance $l_{\text{sides}}$ near the side walls. 
This results in a less uniform macro-scale velocity profile and a smaller slip length. 
% Although the variation of $l_{\text{slip}}$ with $Re_{b}$ remains below 20\% for each geometry considered in this work, this linear relationship is in contrast with the slip length for channels with an array of square cylinders. 
% Here, it was found that $l_{\text{slip}}$ is approximately independent of the Reynolds number (Ref.~\onlinecite{buckinx2022arxiv}). 

Further, we can recognize from Figure \ref{fig:lslip_N2}, as well as Figure \ref{fig:xi_y}, that the shape of the developed macro-scale velocity profile in the side-wall region is nearly independent of the number of unit cells in the lateral direction $N_{2}$. 
Therefore, also the slip length barely varies with $N_{2}$ and the channel aspect ratio. 
For example, for $Re_{b}=28$, $h/l=0.12$, $s/l=0.48$, $t/l=0.02$ and $N_{2} \in (5,17)$, the slip length can be approximated by a constant within a relative error of 3\%: $l_{\text{slip}} / l_{2} \simeq 4.54$. 
This is due to the fact that, for all the considered values of $N_{2}$, the distance over which the side wall influences the flow remains small with respect to the width of a single unit cell $l_{\text{sides}} < l_{2}$. 
Consequently, the spatially periodic flow patterns in the core of the channel, or more precisely $\Omega_{\text{uniform}}$, remain approximately the same for $N_{2} \in (5,17)$, as long as the unit cell geometry remains fixed. 
In turn, also the flow patterns in the side-wall region $\Omega_{\text{sides}}$ will thus remain similar, as the flow in this region is on one side bounded by the flow in $\Omega_{\text{uniform}}$, and must respect the no-slip condition at the other side.
% This is due to the limited extent to which the side wall influences flow with respect to the width of a single unit cell. 
% With respect to the total channel width $N_{2} l_{2}$ with $N_{2} \in (5,17)$, this limited extent results in an insignificant change of the uniform macro-scale flow in $\Omega_{\text{uniform}}$, and hence a similar behavior in the side-wall region $\Omega_{\text{sides}}$. 

\begin{figure}[ht!]
\centering
\begin{minipage}{.475\textwidth}
% \centering
\raggedleft
\includegraphics[scale = 1.0]{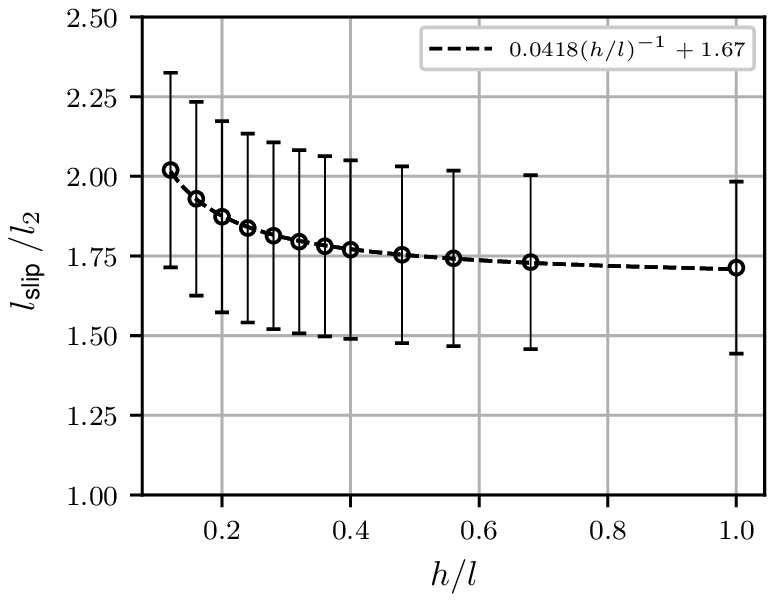}
\caption{\label{fig:lslip_h} Influence of the fin height-to-length ratio on the macro-scale slip length, when $Re_{b} l / (2 L_{3}) = \rho_{f} u_{b} l / \mu_{f} =600$, $N_{1}=20$, $N_{2}=10$, $s/l=0.12$, $t/l=0.02$}
\end{minipage}
\hfill
\begin{minipage}{.475\textwidth}
% \centering
\raggedright
% \raggedleft
\includegraphics[scale = 1.0]{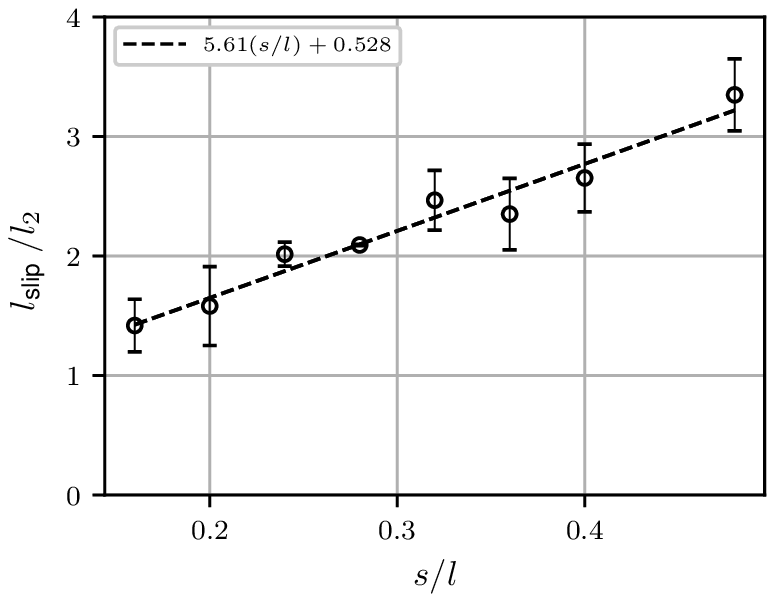}
\caption{\label{fig:lslip_s} Influence of the fin pitch-to-length ratio on the macro-scale slip length, when $Re_{b}=192$, $N_{1}=20$, $N_{2}=10$, $h/l=0.12$, $t/l=0.04$ \newline}
\end{minipage}
\end{figure}

On the contrary, the slip length does depend on the fin height-to-length ratio $h/l$: we observe $l_{\text{slip}} / l_{2} $ to vary linearly with $l/h$. 
To illustrate this, we mention that for the data depicted in Figure \ref{fig:lslip_h}, the correlation $l_{\text{slip}} / l_{2} \simeq 0.0418 (h/l)^{-1} + 1.67$ has an accuracy of 1\%. 
The observation that $l_{\text{slip}}$ increases when the relative fin height $h/l$ decreases, can be explained due to the fact that the lateral velocity $u_{1}$, and especially the transversal velocity $u_{2}$, are dampened when the bottom and top plate become closer to each other. 
As such, the flow in the core of the channel is able to adapt itself over a shorter distance $\Omega_{\text{sides}}$ towards $\Gamma_{\text{sides}}$, in order to satisfy the no-slip condition imposed by the side walls. 
This results in a more uniform macro-scale velocity profile for smaller $h/l$ ratios. 
We note that the proposed correlation between $l_{\text{slip}} / l_{2} $ and  $l/h$ is in line with our expectation that $l_{\text{slip}}$ becomes independent of the channel height for relatively high values of $h/l$, as one finds for arrays of square cylinders (Ref.~\onlinecite{buckinx2022arxiv}). 
Again, this trend is a consequence of the flow patterns becoming more two-dimensional and thus independent of $h/l$ when the fin height increases. 

As shown in Figure \ref{fig:lslip_s}, we also observe a strong (linear) dependence of the macro-scale velocity profile and slip length on the fin pitch-to-length ratio $s/l$.
Specifically, we found that $l_{\text{slip}} / l_{2} \simeq 5.61 (s/l) + 0.528$ for $Re_{b}=192$, $N_{2}=10$, $h/l=0.12$, $t/l=0.04$ and $s/l \in (0.12,0.48)$, at least within a relative error below 8\%. 
Hence, $l_{\text{sides}}$ decreases when the fin pitch $s$ decreases. 
Because a smaller fin pitch $s$ again induces a stronger lateral velocity $u_{2}$, its effect on $l_{\text{sides}}$ is equivalent to that of a higher fin thickness $t/l$.
% This trend further confirms that the slip length increases with the porosity, as observed in (Ref.~\onlinecite{buckinx2022arxiv}). 

From the previous discussion, we conclude that the slip lengths in the side-wall region of micro- and mini-channels with offset strip fins are significantly larger than those in high-porosity arrays of square cylinders for similar boundary conditions (Ref.~\onlinecite{buckinx2022arxiv}). 
This implies that the macro-scale velocity profile in micro- and mini-channels with offset strip fins is relatively more uniform.
In other words, the side walls have a smaller influence on the mass flow rate and macro-scale flow distribution in these channels. 
Finally, we remark that for the same reasons, the common approximation $\boldsymbol{U}_{\text{dev}} \simeq u_b \boldsymbol{e}_1$ is actually quite well justified outside of the side-wall region, according to our DNS results. 
This statement becomes more clear if we inspect the relation between the uniform macro-scale velocity $U$ in the core of the channel and the bulk velocity $u_{b}$, which is given by
\begin{equation}
    U = \frac{N_{2}}{N_{2} - 2 (1-\chi_{u})} u_{b}. 
\end{equation}
Here, the displacement factor $\chi_{u}$ is defined as in (Ref.~\onlinecite{buckinx2022arxiv}), so that it corresponds to the ratio of the mass flow rate through the side-wall region $\Omega_{\text{sides}}$ to the mass flow rate through $\Omega_{\text{uniform}}$, multiplied with the ratio of the cross-sectional area of $\Omega_{\text{uniform}}$ to that of $\Omega_{\text{sides}}$.
As this displacement factor $\chi_{u}$ remains larger than 0.87 for all the Reynolds numbers and channel geometries considered in this work, the ratio $U/u_{b}$ does not exceed 1.04. 

\newpage
\clearpage


\subsection{\label{sec:sidewallmacroclosureforce} Macro-scale closure force in the side-wall region}

Following the work (Ref.~\onlinecite{buckinx2022arxiv}), the developed macro-scale closure force in the side-wall region can be approximated from its developed prediction $ \boldsymbol{b}_{\text{unit}}$: 
% outside the side wall region:
\begin{equation}
\boldsymbol{b} \simeq \xi^{-1} \boldsymbol{b}_{\text{unit}}  \left( U_{\text{dev}}\right),
% b_{1} \simeq \epsilon_{fm} \left( x_{2} \right) \mathrm{\nabla{P}},
\label{eq:closureforcesidewall}
\end{equation}
since 
$\boldsymbol{f}_{\text{closure}} \simeq \boldsymbol{b}$ in $\Omega_{\text{sides}} \cup \Omega_{\text{dev}}$.
We clarify that $ \boldsymbol{b}_{\text{unit}} $ can be evaluated through the developed friction factor correlation (\ref{eq: closure force developed correlations}) and (\ref{eq:friction_correlation}) based on the local macro-scale velocity $U_{\text{dev}} (x_2) $ and the constant array porosity $\epsilon_{f}$. 
% using the local Reynolds number $Re_{l} \left( \boldsymbol{x} \right)$ in $\Omega_{\text{sides}}$. 
Essentially, the approximation (\ref{eq:closureforcesidewall}) implies a balance between the closure force and the spatially constant developed pressure gradient in the side-wall region once the macro-scale flow has become developed:
$\boldsymbol{b} \simeq \mathrm{\nabla{P}} \, \epsilon_{fm} \left( x_{2} \right)$.
Therefore, we can evaluate the developed closure force in the side-wall region directly from the pressure gradient $\mathrm{\nabla{P}}$ which belongs to the equivalent uniform macro-scale velocity $U' =  \xi^{-1} U_{\text{dev}}' $ in $\Omega_{\text{sides}}$. 
% As such, the macro-scale closure force in the side-wall region .

We note that the former approximation for the macro-scale closure force is based on the assumption that the contribution of the momentum dispersion tensor is negligible with respect to the macro-scale closure force at low to moderate Reynolds numbers (Ref.~\onlinecite{buckinx2022arxiv}). 
In addition, we remark that the right-hand side of (\ref{eq:closureforcesidewall}) can be interpreted as a rescaling of the apparent permeability tensor in the side-wall region with the closure variable $\xi$. 
Therefore, there is a close connection to the works of Valdès-Parada and Lasseux (Refs.~\onlinecite{valdes2021novel,valdes2021flow}). 
In these works, a macro-scale flow model is presented for porous media containing a porosity gradient, which resembles a Darcy-type equation with a spatially dependent apparent permeability tensor. 
% the local closure problems for the macro-scale flow in a porous medium near a solid wall, studied in the works of Valdès-Parada and Lasseux (Refs.~\onlinecite{valdes2021novel,valdes2021flow}). 

\begin{figure}[ht]
\includegraphics[scale = 1.0]{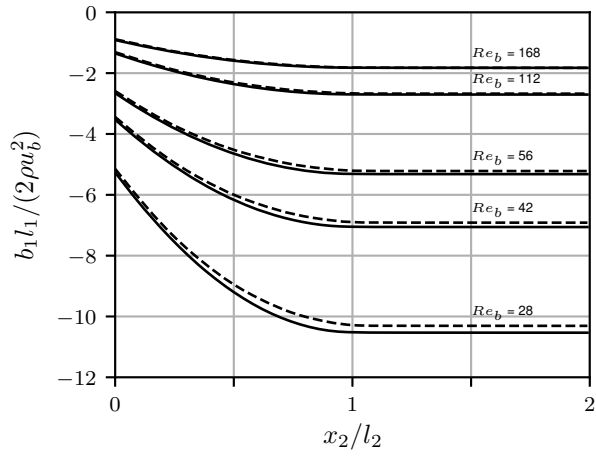}
\caption{\label{fig:b_y_t0.02_h0.12_s0.48} Macro-scale closure force (full) and its prediction (dashed) by the developed correlation from (Ref.~\onlinecite{vangeffelen2021friction}), when $N_{2} = 10$, $h/l=0.12$, $s/l=0.48$, $t/l=0.02$}
\end{figure}

As illustrated in Figure \ref{fig:b_y_t0.02_h0.12_s0.48}, the validity of approximation (\ref{eq:closureforcesidewall}) is supported by our DNS results. 
The agreement between the actual closure force (\ref{eq:closureforce}) and its approximation (\ref{eq:closureforcesidewall}) based on the quadratic velocity profile (\ref{eq:xi}) results in a mean and maximum error of 3\% and 12\% for $b_{1} = \boldsymbol{b} \cdot \boldsymbol{e}_1$, respectively, for all cases regarded in this work. 
Consequently, the preceding knowledge of the slip lengths suffices to reconstruct the closure force $b_{1}$ in the side-wall region (Ref.~\onlinecite{buckinx2022arxiv}). 
% This shows that in the side-wall region $\Omega_{\text{sides}}$, the macro-scale closure force can be reconstructed with the \textit{a priori} knowledge of the macro-scale velocity profile $\xi$, and thus the slip length $l_{\text{slip}}$. 
We remark that since the side-wall region of the channel only extends over the width of one unit cell $l_{2}$, the developed macro-scale velocity profile $\xi$, and thus the slip length $l_{\text{slip}}$, can be computed on a so-called extended unit cell. 
This extended unit cell consists of two adjacent unit cells of which one lies in $\Omega_{\text{sides}}$ and the other in $\Omega_{\text{uniform}}$. 

% In comparison with arrays of less porous fin geometries such as square cylinders (Ref.~\onlinecite{buckinx2022arxiv}), it can be concluded that the slip lengths characterizing the developed macro-scale velocity profiles in $\Omega_{\text{sides}}$ are significantly larger for offset strip fin micro- and mini-channels. 
% This corresponds to a more uniform macro-scale velocity profile and, therefore, a less significant influence of the channel side wall on the mass flow rate distribution and the local apparent permeability tensor. 

% \begin{figure}[ht]
% \centering
% \begin{minipage}{.475\textwidth}
% % \centering
% \raggedleft
% \includegraphics[scale=1.0]{figures/xi_y.eps}
% \caption{\label{fig:xi_y} Macro-scale velocity profile in the side-wall region, when $Re_{b}=600$, $h/l=0.12$, $s/l=0.48$ \newline}
% \end{minipage}
% \hfill
% \begin{minipage}{.475\textwidth}
% % \centering
% \raggedright
% % \raggedleft
% \includegraphics[scale = 1.0]{figures/b_y_t0.02_h0.12_s0.48.eps}
% \caption{\label{fig:b_y_t0.02_h0.12_s0.48} Macro-scale closure force (full) and its prediction (dashed) by the developed correlation from (Ref.~\onlinecite{vangeffelen2021friction}), when $N_{2} = 10$, $h/l=0.12$, $s/l=0.48$, $t/l=0.02$}
% \end{minipage}
% \end{figure}

\newpage
\clearpage

\section{\label{sec:conclusion}Conclusions}

In the present work, the onset of (periodically) developed and quasi-developed flow in offset strip fin micro- and mini-channels has been examined. 
To this end, the complete steady laminar flow fields in various channel geometries were obtained by numerically solving the incompressible Navier-Stokes equations for Reynolds numbers ranging from 28 to 1224. 
It was observed that the onset point of developed flow increases linearly with the Reynolds number and the number of fins along the lateral direction, as well as the fin pitch-to-length ratio.
In addition, this onset point was found to become independent of the fin height for fin height-to-length ratios above one. 
The onset point of quasi-developed flow appears to obey the same scaling laws, although it is almost unaffected by the channel aspect ratio. 
Nevertheless, for all cases considered in this work, the onset point of quasi-developed flow practically coincides with the channel inlet, and the flow development length remains rather small relative to the total channel length. 

Furthermore, we have demonstrated that the macro-scale flow in these channels can be treated as entirely developed, by numerically computing the macro-scale velocity field and macro-scale pressure gradient through a double volume-averaging operation. 
The developing macro-scale velocity profile near the channel inlet was found to deviate only modestly from that in the developed region, due to the small flow development length. 
In particular, we found the angle of attack of the macro-scale flow to stay below 1$^{\circ}$ over the entire channel. 
Also the macro-scale pressure gradient in the developing region was found to remain virtually constant. 
As a result, its local values throughout the entire channel are well captured by the developed friction factor correlation from our previous work, which yields a mean and maximum error of 2\% and 15\% for the illustrated cases. 
Therefore, also the overall pressure drop over the entire channel can be accurately predicted by this developed friction factor correlation. 

Moreover, we conclude that the macro-scale flow in the developing flow region is essentially quasi-developed, so that its main features are characterized by a single exponential mode. 
The amplitude and shape of this mode seem to be marginally affected by the Reynolds numbers and the channel aspect ratio, whereas the mode eigenvalue varies clearly inversely linear with both the Reynolds number and channel width. 
Consequently, the eigenvalues of the quasi-developed flow are directly responsible for the observed scaling laws for the onset point of developed flow with the former parameters. 
In general, the mode amplitudes are much smaller than those observed in high-porosity arrays of square cylinders, while the corresponding eigenvalues are significantly larger. 
This again explains the relatively rapid flow development in micro- and mini-channels with offset strip fins. 

Finally, we conclude that the developed macro-scale velocity profile in the side-wall region of the channel has approximately a quadratic shape. 
This shape is described by a single slip length, which varies linearly with the Reynolds number and the fin pitch-to-length ratio. 
Yet, the shape profile is independent of the aspect ratio and barely affected by the channel height. 
If one re-scales the developed friction factor in the developed region with this shape profile, one obtains an approximation for the macro-scale pressure gradient in the side-wall region, which results in a typical mean and maximum error of less than 5\% and 15\%, respectively.

\newpage
\clearpage

\section{\label{sec:contributions}Contributions}
The macro-scale model based on the double volume-averaging operation and its computational framework were developed and validated by G. Buckinx. 
All flow simulations, as well as post-processing calculations, were carried out by A. Vangeffelen. 
The interpretation of the results was done by A. Vangeffelen, with the input from G. Buckinx regarding the available literature. 
A. Vangeffelen and G. Buckinx wrote the paper with the input from C. De Servi, M. R. Vetrano and M. Baelmans.

\section{\label{sec:acknowledgements}Acknowledgements}
The work presented in this paper was partly funded by the Research Foundation — Flanders (FWO) through the post-doctoral project grant 12Y2919N of G. Buckinx, and partly by the Flemish Institute for Technological Research (VITO) through the Ph.D. grant 1810603 of A. Vangeffelen. 
The VSC (Flemish Supercomputer Center), funded by the Research Foundation - Flanders (FWO) and the Flemish Government, provided the resources and services used in this work.

\newpage
\clearpage

\appendix

\newpage
\clearpage

\nocite{*}
\bibliography{aipsamp}% Produces the bibliography via BibTeX.

\end{document}